\documentclass[aps, amssymb, amsmath, nofootinbib, showpacs, notitlepage, byrevtex, showkeys, prd, %preprint%,tightenlines%
]{revtex4-1}

\usepackage{graphicx}
\usepackage{subfigure}
\usepackage{bm}
\usepackage{color}
\usepackage{framed}

\renewcommand{\vec}[1]{\boldsymbol{#1}}

\newcommand{\beq}{\begin{equation}}
\newcommand{\eeq}{\end{equation}}

\newcommand{\ii}{\mathrm{i}}
\newcommand{\dd}{\mathrm{d}}
\newcommand*{\da}[1][]{\mathop{\mathrm{d}\mkern-7mu\mathchar'26\mkern-1mu^{#1}}\mkern-4mu}

\newcommand{\tr}{\mathrm{tr}}
\newcommand{\Ncc}{N_{\mathrm{C}}}
\newcommand{\Det}{\mathrm{det}}

\newcommand{\vx}{\vec{x}}
\newcommand{\vy}{\vec{y}}
\newcommand{\vz}{\vec{z}}
\newcommand{\vp}{\vec{p}}
\newcommand{\vq}{\vec{q}}

\newcommand{\vk}{\vec{k}}
\newcommand{\vA}{\vec{A}}
\newcommand{\vB}{\vec{B}}
\newcommand{\vD}{\vec{D}}
\newcommand{\vPi}{\mathbf{\Pi}}
\newcommand{\valpha}{\boldsymbol{\alpha}}

\newcommand{\ve}{\vec{e}}

\newcommand{\nf}{\mathfrak{n}}
\newcommand{\mf}{\mathfrak{m}}
\newcommand{\lf}{\mathfrak{l}}
\newcommand{\he}{\hat{\ve}}

\usepackage[font=small]{caption}
\begin{document}

\title{Chiral symmetry restoration at finite temperature within the Hamiltonian approach to QCD in Coulomb gauge}

\author{Markus Quandt, Ehsan Ebadati, Hugo Reinhardt and Peter Vastag}
\affiliation{Institut f\"ur Theoretische Physik\\
Auf der Morgenstelle 14\\
D-72076 T\"ubingen\\
Germany}
\date{\today}

\begin{abstract}

The chiral phase transition of the quark sector of QCD is investigated within the Hamiltonian approach in Coulomb gauge. Finite temperature $T$ is introduced by compactifying one spatial dimension, which makes all thermodynamical quantities accessible from the ground state on the spatial manifold $\mathbb{R}^2 \times S^1(1/T)$. Neglecting the coupling between quarks and transversal gluons, the equations of motion of the quark sector are solved numerically and the chiral quark condensate is evaluated and compared to the results of the usual canonical approach to finite-temperature Hamiltonian QCD based on the density operator of the grand canonical ensemble. For zero bare quark masses, we find a second-order chiral phase transition with a critical temperature of about $92 \, \mathrm{MeV}$. If the Coulomb string tension is adjusted to reproduce the 
phenomenological value of the quark condensate, the critical temperature increases to
$118\,\mathrm{MeV}$.

\end{abstract}
\maketitle

\section{Introduction} \label{Abschn: Einleitung}

Understanding the phase diagram of quantum chromodynamics (QCD) is still one of the most challenging problems in particle physics \cite{Karsch2002, Fukushima2011}. Lattice calculations can shed some light on its structure for vanishing baryon density but still suffer from the so-called sign problem in the general case of finite densities \cite{Karsch2002, Gattringer2016}. To overcome this problem, in the past two decades several non-perturbative continuum approaches, which do not suffer from the sign problem, have been developed \cite{Fischer2006, *Alkofer2001, *Binosi2009, *Pawlowski2007, *Gies2012, *Quandt2014, *Quandt2015}, one of them being the variational approach to Hamiltonian QCD in Coulomb gauge \cite{Feuchter2004, *Feuchter2005}, see ref.~\cite{Reinhardt2017} for a recent review.

In ref.~\cite{RV2016}, the dressed Polyakov loop, the order parameter for confinement, and the chiral quark condensate, the order parameter for the spontaneous breaking of chiral symmetry, have been evaluated within this approach for vanishing chemical potential (i.e.~baryon density). Thereby, finite temperatures were introduced by compactifying one spatial dimension using the alternative formulation of finite-temperature Hamiltonian quantum field theory proposed in ref.~\cite{Reinhardt2016}. While the pseudo-critical temperatures of the chiral and, respectively, deconfinement phase transition were in good agreement with lattice data, the width of the transition region and the order of the chiral phase transition turned out to be at odds with the lattice predictions. This was suspected to be correlated to the neglect of the temperature dependence of the quark and gluon propagator, which were replaced by their zero-temperature limits to avoid the numerically highly expensive solution of the finite-temperature equations of motion.

In the present paper, we solve the quark part of these equations numerically. Thereby, we ignore the coupling of the quarks to the (transversal) spatial gluons. This corresponds to a confining quark model -- the so-called Adler--Davis model \cite{Adler1984} -- which was considered in refs.~\cite{Davis1984, Kocic1986, Alkofer1989, LS2010} in the standard canonical formulation of finite-temperature quantum field theories. From our solution, we calculate the chiral condensate and compare it with the result of previous work.

The organization of the rest of this paper is as follows: In section \ref{Abschn: HamiltonscheFormulierung}, we briefly review the essential ingredients of the novel approach to finite-temperature Hamiltonian quantum field theory developed in ref.~\cite{Reinhardt2016} and its application to QCD in Coulomb gauge given in ref.~\cite{RV2016}. The numerical solution of the quark equations of motion is described in detail 
in section \ref{Abschn: Numerik}. The results for the mass function and the chiral condensate are 
presented in section \ref{sec:results}, and we conclude the manuscript with a brief summary, 
some comments and an outlook on future directions in section \ref{sec:conclusions}.

\section{The quark sector of finite-temperature QCD} \label{Abschn: HamiltonscheFormulierung}

Below, we briefly discuss the main ingredients of the Hamiltonian approach to the quark sector of QCD when finite temperatures are introduced by compactifying a spatial dimension, for which we choose w.l.o.g.~the 3-axis. For a more detailed description and a discussion of full QCD, the interested reader may consult refs.~\cite{Reinhardt2016, RH2013, RV2016}.

Let $H$ be the QCD Hamiltonian in Coulomb and Weyl gauge on the compactified spatial manifold $\mathbb{R}^2 \times S^1(\beta)$, where $\beta = 1/T$ denotes the inverse temperature. One can then show \cite{Reinhardt2016} that the grand canonical partition function at finite temperature $T$ and chemical potential $\mu$ is given by
\beq
\mathcal{Z} = \lim_{\ell \to \infty} \, \exp\Bigl(-\ell E_0(\beta, \mu)\Bigr) \, , \label{Gl: Zustandssumme}
\eeq
where $\ell \to \infty$ is the length of the uncompactified spatial dimensions and $E_0$ is the smallest eigenvalue of the pseudo-Hamiltonian
\beq
\widetilde{H}(\beta, \mu) \equiv H + \ii \mu \int_{\beta} \dd^3 x \, {\psi^m}^{\dagger}(\vx) \alpha_3 \psi^m(\vx) \, . \label{Gl: PseudoHamiltonian}
\eeq
Here, $\alpha_i$ denotes the usual Dirac matrices and $\psi$ is 
the quark field which has to fulfill the anti-periodic boundary condition
\beq
\psi^m(x_1, x_2, x_3 = \beta/2) = -\psi^m(x_1, x_2, x_3 = -\beta/2)
\eeq
on the compactified manifold, while for the bosonic fields $A$ the periodic condition
\beq
\vA^a(x_1, x_2, x_3 = \beta/2) = \vA^a(x_1, x_2, x_3 = -\beta/2)
\eeq
holds ($m$ and $a$ are color indices in the fundamental and  adjoint, respectively, 
representation). Furthermore, we have introduced the short-hand notation
\beq
\int_{\beta} \dd^3 x \equiv \int \dd x_1 \int \dd x_2 \int_{-\beta/2}^{\beta/2} \dd x_3
\eeq
for the spatial integration.

Let us stress that the novel finite-temperature Hamiltonian approach proposed in ref.~\cite{Reinhardt2016} and leading to eq.~(\ref{Gl: Zustandssumme}) is equivalent to the familiar finite-temperature (imaginary-time) approach for any $O(4)$ (i.e.~relativistic) invariant quantum field theory. It is, however, advantageous in a Hamiltonian formulation in the sense that it does not require to explicitly carry out the thermal expectation values with the grand canonical density operator
\beq
\varrho = \exp\bigl(-\beta [\mathcal{H} - \mu N]\bigr) \label{Gl: DichteOperator}
\eeq
(with $\mathcal{H}$ being the Hamiltonian on $\mathbb{R}^3$ and $N$ being the fermionic particle-number operator) over the whole Fock space. Rather the thermal quantities like the partition function (\ref{Gl: Zustandssumme}) are obtained from the vacuum state on $\mathbb{R}^2 \times S^1(\beta)$ alone (see below). Thus, the novel approach avoids introducing additional approximations to the Hamiltonian in the grand canonical density operator $\exp\bigl(-\beta [\mathcal{H} - \mu N]\bigr)$, which is certainly an advantage. The novel approach is not manifestly spatial $O(3)$-invariant in the same way as the standard Hamiltonian approach based on canonical quantization is not manifestly Lorentz-invariant. However, the $O(3)$ and $O(4)$ invariance is hidden and recovered when the approach is carried out exactly. Let us also mention that in the novel approach the $T \to 0$ limit can be easily taken after Poisson resummation, see Refs.~\cite{Reinhardt2016} 
for more details. This fact will be exploited in the discussion following eq.~(\ref{gap0}) below.

The QCD Hamiltonian $H$ entering eq.~(\ref{Gl: PseudoHamiltonian}) is given by \cite{Christ1980}
\beq
H = H_{\mathrm{D}} + H_{\mathrm{YM}} + H_{\mathrm{C}} \, , \label{Gl: QCDHamiltonian}
\eeq
where
\begin{subequations} \label{Gl: DiracHamiltonian}
\begin{align}
H_{\mathrm{D}} &= H_{\mathrm{D}}^0 + H_{\mathrm{D}}^A \, , \\
H_{\mathrm{D}}^0 &= \int_{\beta} \dd^3 x \, {\psi^m}^{\dagger}(\vx) \Bigl[-\ii \valpha \cdot \nabla + \gamma_0 m_{\mathrm{Q}}\Bigr] \psi^m(\vx) \, , \\
H_{\mathrm{D}}^A &= g \int_{\beta} \dd^3 x \, {\psi^m}^{\dagger}(\vx) \valpha \cdot \vA^a(\vx) t_a^{m n} \psi^n(\vx)
\end{align}
\end{subequations}
is the quark single-particle Dirac Hamiltonian with $g$ being the strong coupling constant, $m_{\mathrm{Q}}$ the bare quark mass, $\gamma_0$ the usual Dirac matrix and $t_a$ the color generator in the fundamental representation. The second term in eq.~(\ref{Gl: QCDHamiltonian}) is the gluonic Yang--Mills Hamiltonian
\beq
H_{\mathrm{YM}} =  \frac{1}{2} \int_{\beta} \dd^3 x \, J^{-1}[A] \Pi_i^a(\vx) J[A] \Pi_i^a(\vx) + \frac{1}{2} \int_{\beta} \dd^3 x \, \vB^a(\vx) \cdot \vB^a(\vx) \, ,
\eeq
where $\Pi = -\ii \delta / \delta A$ is the canonical momentum operator (which agrees with the color electric field),
\beq
\vB^a = \nabla \times \vA^a - \frac{1}{2} g f^{a b c} \vA^b \times \vA^c
\eeq
is the color magnetic field and
\beq
J[A] = \Det\bigl(\hat{G}^{-1}\bigr) \, , \quad \bigl(\hat{G}^{-1}\bigr)^{a b}(\vx, \vy) \equiv \bigl(-\nabla \cdot \hat{\vD}\bigr)^{a b}(\vx, \vy)
\eeq
denotes the Faddeev--Popov determinant with
\beq
\hat{\vD}^{a b}(\vx) = \delta^{a b} \nabla - g f^{a c b} \vA^c(\vx)
\eeq
being the covariant derivative in the adjoint representation. Finally, 
\beq
H_{\mathrm{C}} = \frac{g^2}{2} \int_{\beta} \dd^3 x \int_{\beta} \dd^3 y \, J^{- 1}[A] \rho^a(\vx) J[A] \hat{F}^{ab}(\vx, \vy) \rho^b(\vy) \label{Gl: Coulombterm}
\eeq
is the so-called color Coulomb interaction which contains, besides the color density
\beq
\rho^a(\vx) = \rho_{\mathrm{Q}}^a(\vx) + \rho_{\mathrm{YM}}^a(\vx) = \psi^{\dagger}(\vx) t_a \psi(\vx) + f^{abc} \vA^b(\vx) \cdot \vPi^c(\vx) \label{Gl: Ladungsdichte}
\eeq
of quarks and gluons, the non-Abelian Coulomb kernel
\beq
\hat{F}^{a b}(\vx, \vy) = \int_{\beta} \dd^3 z \, \hat{G}^{a c}(\vx, \vz) (-\Delta_{\vz}) \hat{G}^{c b}(\vz, \vy) \, .
\eeq

From eq.~(\ref{Gl: Zustandssumme}) it follows that all thermodynamical quantities can be obtained from the ground state $\vert \phi \rangle$ of the pseudo-Hamiltonian $\widetilde{H}$ which fulfills the functional Schr\"odinger equation $\widetilde{H} \vert \phi \rangle = E_0 \vert \phi \rangle$ \cite{Reinhardt2016}. Solving the functional Schr\"odinger equation is, thus, the aim of the Hamiltonian approach. On the compactified manifold $\mathbb{R}^2 \times S^1(\beta)$, this has been first tackled in ref.~\cite{Heffner2015} for the Yang--Mills sector and was recently extended to full QCD in ref.~\cite{RV2016}. Thereby, the ground state was calculated in an approximative way by using the variational principle: 
Using Gaussian type ans\"atze for both the bosonic and fermionic\footnote{The fermionic part of the vacuum wave functional in ref.~\cite{RV2016} includes also the coupling of the quarks to the transversal gluons and is hence not strictly Gaussian.} parts of the vacuum wave functional $\vert \phi \rangle$, the expectation value $\langle \phi \vert \widetilde{H} \vert \phi \rangle$ was calculated on two-loop level. From its minimization, a set of coupled integral equations for the variational kernels contained in the ansatz for the wave functional $\vert \phi \rangle$ was obtained. While the so-called gap equation for the Yang--Mills sector was solved numerically in ref.~\cite{Heffner2015}, the full coupled equations were left unsolved in ref.~\cite{RV2016} due to the high numerical expense. Instead, the zero-temperature propagators obtained in ref.~\cite{QCDT0Rev} were used to calculate the dressed Polyakov loop and the temperature dependence of the chiral quark condensate for $\mu = 0$. Remarkably, within these approximations the inclusion of the coupling of the quarks to the transverse spatial gluons showed only a negligible effect on the pseudo-critical temperatures of the deconfinement and chiral phase transitions.

\medskip
In the present paper, we will give the numerical solution of the finite-temperature variational
equations of motion for the quark sector and calculate the chiral condensate from it. Since the
numerical cost is substantially higher for solving the full coupled equations, we will thereby
neglect the coupling between quarks and transversal gluons.\footnote{In ref.~\cite{RV2016}, this 
case was labeled as $g = 0$ limit, although the coupling of the quarks to the temporal 
vector field $A_0$ and hence the Coulomb interaction is retained.} Although it is not clear whether
the effect of the coupling of the quarks to the transversal spatial gluons is still subleading at
finite-temperature, this will enable us to study the effects of the temperature-dependence of the
solution on the order and width of the chiral phase transition. Furthermore, it also allows for
comparison between the compactified theory and the usual grand canonical approach to finite
temperatures in Hamiltonian QCD considered in refs.~\cite{Davis1984, Kocic1986, Alkofer1989, LS2010}.

Neglecting the coupling between quarks and transverse gluons, the fermionic part of the QCD Hamiltonian\footnote{For a discussion of the Yang--Mills part see ref.~\cite{Heffner2015}.} reduces to
\beq
H_{\mathrm{Q}} = H_{\mathrm{D}}^0 + H_{\mathrm{C}}^{\mathrm{Q Q}} \, , \label{Gl: QuarkHamiltonian}
\eeq
where $H_{\mathrm{D}}^0$ [Eq.~(\ref{Gl: DiracHamiltonian})] is the free Dirac Hamiltonian and $H_{\mathrm{C}}^{\mathrm{Q Q}}$ follows from the Coulomb term (\ref{Gl: Coulombterm}) after substituting $\rho \to \rho_{\mathrm{Q}}$ [Eq.~(\ref{Gl: Ladungsdichte})]. Note that this implies the cancellation of the Faddeev--Popov determinant in eq.~(\ref{Gl: Coulombterm}). Furthermore, on two-loop level the non-Abelian Coulomb kernel can be replaced by its (Yang--Mills) vacuum expectation value,
\beq
g^2 \langle \hat{F}^{a b}(\vx, \vy) \rangle_{\mathrm{YM}} \approx \delta^{a b} V_{\mathrm{C}}(|\vx - \vy|) \, ,
\label{cpot}
\eeq
which plays the role of a confining quark potential, $V_{\mathrm{C}}(|\vx - \vy|) =
\sigma_{\mathrm{C}} |\vx - \vy|$ at $|\vx-\vy|\to\infty$, where $\sigma_{\mathrm{C}}$ is the
Coulomb string tension \cite{ERS2007}.

Neglecting the coupling between quarks and transversal gluons, the ansatz for the fermionic part of the vacuum wave functional from ref.~\cite{RV2016} reduces to the BCS-type functional
\beq
\vert \phi \rangle = \exp\left(-\int_{\beta} \dd^3 x \int_{\beta} \dd^3 y \, {\psi_+^m}^{\dagger}(\vx) \gamma_0 S(\vx - \vy) \psi_-^m(\vy)\right) \vert 0 \rangle \, , \label{Gl: Ansatz}
\eeq
where $S$ is a scalar variational kernel, $\psi_{\pm}$ denotes the positive/negative spectral projection of the quark field $\psi$ and $\vert 0 \rangle$ is the bare vacuum of the Dirac sea, fulfilling $\psi_+ \vert 0 \rangle = \psi_-^{\dagger} \vert 0 \rangle = 0$. This type of ansatz together with the Hamiltonian (\ref{Gl: QuarkHamiltonian}) corresponds to the confining quark model (Adler--Davis model) considered e.g.~in refs.~\cite{FM1982, LeYaouanc1984, Adler1984, AA1988} at zero temperature and in refs.~\cite{Davis1984, Kocic1986, Alkofer1986, Alkofer1989, LS2010} in the usual canonical approach to finite temperatures and densities. For explicit calculations, it is convenient to switch to the momentum 
space representation using
\beq
S(\vx) = \int_{\beta} \da^3 p \, \exp\bigl(\ii (\vp_{\perp} + \Omega_{\nf} \he_3) \cdot \vx\bigr) S(\vp_{\perp}, \Omega_{\nf}) \, ,
\eeq
where $\vp_{\perp} = p_1 \he_1 + p_2 \he_2$ is the planar momentum 
and 
\beq
\Omega_{\nf} = \frac{2 \nf + 1}{\beta} \pi\,,\qquad\qquad \nf \in \mathbb{Z}
\eeq
are the fermionic Matsubara frequencies resulting from the Fourier transformation of the (compactified) spatial component $x_3$. Furthermore, we have introduced the short-hand notation [$\da = \dd / (2 \pi)$]
\beq
\int_{\beta} \da^3 p \equiv \int \da^2 p_{\perp} \, \frac{1}{\beta} \sum_{\nf = -\infty}^{\infty} \, .
\label{matsmeasure}
\eeq
In the following, we focus on the limit of vanishing chemical potential
($\mu = 0$) and chiral quarks ($m_{\mathrm{Q}} = 0$). From the variational principle $\langle \phi \vert H_{\mathrm{Q}} \vert \phi \rangle \to \min$ one finds then the following integral equation for the variational kernel $S$
\begin{align}
\sqrt{k_{\perp}^2 + \Omega_{\lf}^2}\, S(\vk_{\perp}, \Omega_{\lf}) &= \frac{C_{\mathrm{F}}}{2} \int_{\beta} \da^3 p \, V_{\mathrm{C}}\bigl(\vp_{\perp} - \vk_{\perp} + (\Omega_{\nf} - \Omega_{\lf}) \he_3\bigr) \frac{1}{1 + S^2(\vp_{\perp}, \Omega_{\nf})} \nonumber \\
&\phantom{=}\,\, \times \left[S(\vp_{\perp}, \Omega_{\nf}) \Bigl(1 - S^2(\vk_{\perp}, \Omega_{\lf})\Bigr) \vphantom{\frac{\vp_{\perp} + 
\Omega_{\nf} \he_3}{\sqrt{p_{\perp}^2 + \Omega_{\nf}^2}} \cdot \frac{\vk_{\perp} + \Omega_{\lf} \he_3}{\sqrt{k_{\perp}^2 + 
\Omega_{\lf}^2}}} - S(\vk_{\perp}, \Omega_{\lf}) \Bigl(1 - S^2(\vp_{\perp}, \Omega_{\nf})\Bigr) \frac{\vp_{\perp} + 
\Omega_{\nf} \he_3}{\sqrt{p_{\perp}^2 + \Omega_{\nf}^2}} \cdot \frac{\vk_{\perp} + 
\Omega_{\lf} \he_3}{\sqrt{k_{\perp}^2 + \Omega_{\lf}^2}}\right] , \label{Gl: GapgleichungS}
\end{align}
where $C_{\mathrm{F}} = \frac{\Ncc^2 - 1}{2 \Ncc}$ is the value of the quadratic Casimir of the color
group $SU(\Ncc)$ \cite{RV2016} and $p_{\perp} = |\vp_{\perp}|$. For the numerical solution it is,
however, more convenient to rewrite the scalar kernel $S$ in terms of the effective quark mass 
function
\beq
M(\vp_{\perp}, \Omega_{\nf}) = \frac{2 \sqrt{p_{\perp}^2 + \Omega_{\nf}^2} \,S(\vp_{\perp}, \Omega_{\nf})}{1 - S^2(\vp_{\perp}, 
\Omega_{\nf})} \label{Gl: Massenfunktion}
\eeq
which transforms the gap equation (\ref{Gl: GapgleichungS}) to
\beq
M(\vk_{\perp}, \Omega_{\lf}) = \frac{C_{\mathrm{F}}}{2} \int_{\beta} \da^3 p \,
V_{\mathrm{C}}\bigl(\vp_{\perp} - \vk_{\perp} + (\Omega_{\nf} - \Omega_{\lf}) \he_3\bigr)
 \frac{M(\vp_{\perp}, \Omega_{\nf}) - M(\vk_{\perp}, \Omega_{\lf}) \frac{\vp_{\perp} \cdot \vk_{\perp} +
\Omega_{\nf} \Omega_{\lf}}{k_{\perp}^2 + \Omega_{\lf}^2}}{\sqrt{p_{\perp}^2 + \Omega_{\nf}^2 +
 M^2(\vp_{\perp}, \Omega_{\nf})}} \, . 
\label{gapeq}
\eeq
Assuming the linearly rising form\footnote{In a dynamical calcluation \cite{ERS2007}, 
one find a potential $V_C(|\vx|)$ which can be nicely fitted by a linearly rising 
term $\sim|\vx|$ plus an ordinary Coulomb term $\sim 1/|\vx|$. The latter is also found 
in perturbation theory \cite{Campagnari2009}, but neglected in the present paper since 
it is infrared suppressed.} 
$V_\mathrm{C}(\mathbf{x}) = \sigma_\mathrm{C}\,|\mathbf{x}| $ for the non-Abelian Coulomb potential
(\ref{cpot}), its Fourier transform in the gap equation is given by 
\begin{align}
V_\mathrm{C}(\mathbf{p}) = \frac{8 \pi \sigma_\mathrm{C}}{|\mathbf{p}|^4}\,.
\label{Vc}
\end{align} 
The Coulomb string tension $ \sigma_{\mathrm{C}}$ entering this expression sets the overall 
scale in the present model. Lattice and continuum calculations 
\cite{Nakagawa:2006fk,Greensite:2014bua,Golterman:2012dx,Burgio:2015hsa,Burgio:2016nad}
favour values $\sigma_\mathrm{C} / \sigma \approx 2\ldots 4$ in terms 
of the Wilson string tension $\sigma$, with the rather large uncertainties coming from the 
extrapolation of the lattice Coulomb potential in the deep infrared. With the standard 
value $\sigma = (440 \mathrm{MeV})^2$ for the Wilson string tension, this puts 
$\sqrt{\sigma_\mathrm{C}}$ in the range $650\,\mathrm{MeV}\ldots 880\,\mathrm{MeV}$.
In the present work, we will use a standard value of 
$\sqrt{\sigma_\mathrm{C}} \approx 700\,\mathrm{MeV}$
corresponding to $\sigma_\mathrm{C} / \sigma \approx 2.5 $, 
but we should be aware that this stipulation easily has uncertainties of up to $20 \%$. 

\medskip\noindent
For a numerical evaluation, eq.~(\ref{gapeq}) is not directly useful, since the entire 
calculation is dominated by the pole of the Coulomb potential for the single 
frequency $\Omega_\nf = \Omega_\lf$ which -- in constrast to the $T=0$ equation 
discussed below -- is not lifted by the integration measure. We thus have to 
introduce a small mass parameter $\mu$ (not to be confused with the chemical potential)
to regularize the potential, and the entire 
calculation becomes very sensitive to the actual value of $\mu$ and the number 
of Matsubara frequencies included in the numerical code. We will present more details
on the Matsubara type of gap equation (\ref{gapeq}) in section \ref{sec:results},
but follow, for the main part of this paper, we follow a different route which also 
brings the underlying physics to the fore, viz.~we \emph{Poisson resum} the Matsubara 
series. For fermions, this is based on the simple distributional identity
\beq
\frac{1}{\beta} \sum_{\nf \in \mathbb{Z}} f\Big(\frac{2\nf+1}{\beta}\,\pi\Big) = 
\sum_{\mf \in \mathbb{Z}} \int\limits_{-\infty}^\infty \da p_z \,
f(p_z)\,(-1)^\mf\,\exp(\ii \mf \beta p_z)
\eeq
valid for a suitable test function $f$. If we use this equation to replace the Matsubara 
sum we find, after combining  terms, shifting the loop momentum $\vp \to \vq \equiv \vp - \vk$, 
and moving the Poisson sum outermost:
\begin{equation}
M(\vk) = 4 \pi \sum_{\mf=-\infty}^\infty (-1)^\mf\,\int  \da^3 q
\cos(\beta \mf (q_z + k_z))\,V(|\vq|) \,\frac{M(\vq + \vk) - 
\big[1 + \frac{\vk\cdot\vq}{\vk^2}\big]\,M(\vk)}
{\sqrt{(\vq + \vk)^2 + M(\vq + \vk)^2}}\,.
\label{gap0}
\end{equation} 
For any index $\mf$, the integral under the sum is bound by the $\mf=0$ contribution,
i.e.~the $T=0$ limit. The zero temperature equation is, however, known to be both 
ultraviolet and infrared finite \cite{RV2016, QCDT0Rev} and the same must hence 
hold for each integral in the Poisson sum (\ref{gap0}) separately. (We will corroborate
this assertion further below.) The infrared singularity of the Matsubara formulation 
now re-appears as convergence issue of the Poisson sum but, as we will demonstrate
below, this issue can be handled analytically.

To close this section, we note that the ansatz (\ref{Gl: Ansatz}) leads to 
the following expression for the chiral quark condensate:
\begin{align}
\langle \bar{\psi} \psi \rangle = 
\langle \phi \vert {\psi}^{\dagger}(\vx) \gamma_0 \psi(\vx) \vert \phi \rangle &= -2 \Ncc
\int_{\beta} \da^3 p \, \frac{M(\vp_{\perp}, \Omega_{\nf})}{\sqrt{p_{\perp}^2 + \Omega_{\nf}^2 +
M^2(\vp_{\perp}, \Omega_{\nf})}}
\nonumber \\[2mm]
&= - 2 \Ncc \sum_{\mf=-\infty}^\infty (-1)^\mf  \int \frac{d^3 p}{(2\pi)^3}
\cos(\mf \beta p_z)\,\frac{M(\vp)}{\sqrt{\vp^2 + M(\vp)^2}}\,.
\label{condensate}
\end{align}

\section{Numerical Method} \label{Abschn: Numerik}
In this section, we sketch the numerical techniques necessary to solve
eq.~(\ref{gap0}). To fix our notation and discuss some numerical optimiziation, 
we briefly revisit the $T=0$ case.

\subsection{The zero temperature case revisited}
\label{sec:temp0}
The zero-temperature gap equation is simply the $\mf = 0$ contribution from 
eq.~(\ref{gap0}). To study it numerically, we measure all dimensionfull 
quantities in units of the mass scale 
\begin{align}
m_0^2 = C_{\mathrm{F}}\,\sigma_{\mathrm{C}} \approx (800\,\mathrm{MeV})^2\qquad\qquad
\text{for\quad} G= SU(3)\,.
\label{scale}
\end{align}
As explained earlier, this stipulation has rather large uncertainties from the 
lattice calculations of $\sigma_\mathrm{C}$ so will have all absolute numbers 
quoted in the present work. In the discussion, we will also present results 
for the quark condensate and the critical temperature, when $m_0$ is adjusted
to match the lattice findings for the condensate at $T=0$.

Next, we introduce spherical coordinates and exploit the rotational symmetry 
of the $T=0$ system to eliminate the azimuthal angle. This gives
\begin{align}
M(k) = \frac{1}{\pi}\int\limits_0^\infty \dd q\,\int\limits_{-1}^1 \dd\xi\,q^2\,U(q)\frac{
M(Q) - \big[1 + q \xi / k\big]\,M(k)}{\sqrt{Q^2 + M(Q)^2}}\,,
\label{zero}
\end{align}
where 
\beq
Q \equiv |\vk + \vq | = \sqrt{k^2 + q^2 + 2 k q \xi}\,,\qquad\qquad\quad
\xi \equiv \cos \sphericalangle(\vk, \vq),
\eeq
and we have indicated that the scalar mass function can only depend on 
$k = |\vk|$ due to spherical symmetry. The prefactors in the mass scale 
eq.~(\ref{scale}) were chosen such that all clutter is removed from the 
Coulomb potential, which now simply reads 
\begin{align}
U(q) = \frac{1}{q^4}\,.
\end{align}
It is easy to see that the momentum integral in eq.~(\ref{zero}) is 
ultraviolet convergent as long as $M(k)$ is bounded at $k\to \infty$. 
In the infrared, the superficial $1/q$ pole in the integrand disappears 
after integration over $\xi$, and the equation is infrared finite as well.
However, solving eq.~(\ref{zero}) by iteration is very unstable and requires 
substantial underrelaxation  for convergence. 
As a consequence, a huge number of iterations (up to 20,000) is necessary 
to find the solution with high accuracy. To better understand this 
behaviour, it is convenient to rewrite eq.~(\ref{zero}) in 
\emph{quotient form} by collecting all pieces that contain the mass 
as a function of the \emph{external} momentum,
\begin{align}
M(k) = \frac{\displaystyle\frac{1}{\pi}\int_0^\infty \dd q\,\int_{-1}^1 \dd\xi
\,q^2\,U(q)\frac{M(Q)}{\sqrt{Q^2 + M(Q)^2}}}
{1 + \displaystyle \frac{1}{\pi}\int_0^\infty \dd q\,\int_{-1}^1 \dd\xi
\,q^2\,U(q)\frac{1 + q \xi / k}{\sqrt{Q^2 + M(Q)^2}}}\,.
\label{quotzero}
\end{align}
At first sight, this seems like a very bad way to rewrite the equation, 
since both the numerator and denominator are now infrared \emph{divergent}.
If we regularize the divergence by a lower cutoff $\mu$ to the momentum
integral, it is easy to see that the leading $\mu\to 0$ contributions
in numerator and denominator differ only by a factor $M(k)$. 
Thus, the r.h.s.~of eq.~(\ref{quotzero}) reads, at small 
regulators,  
\beq
M(k) + \mu \,\left\{\frac{2}{\pi}\,\frac{M(k)}{\sqrt{k^2 + M(k)^2}}
+ \text{non-local}\right\} + \mathcal{O}(\mu^2)\,.
\eeq
Iterating eq.~(\ref{quotzero}) therefore produces changes to the mass 
function which are of order $\mathcal{O}(\mu) $, i.e.~the small infrared 
regulator also limits the speed at which the iteration progresses. This 
is what gives eq.~(\ref{quotzero}) its inherent stability: any form of 
the gap equation in which the integrals are infrared \emph{finite} will produce 
iteration changes of order $\mathcal{O}(1)$ which are way too large 
and must hence be substantially underrelaxed. By contrast, eq.~(\ref{quotzero})
can be overrelaxed  without loosing stability. 

As often, stability does not automatically imply efficiency: since 
eq.~(\ref{quotzero}) makes very small progress in each step, a large 
number of more than 7,000 iterations is still necessary to solve it. This can, 
however, be cured by using \emph{sequence accelerators} for the iteration,
which improve convergence speed without sacrifying stability. In the 
present case, we have tested both a variant of Aitken's $\Delta^2$-process 
\cite{Aitken:1926} and Anderson's higher degree secant method 
\cite{Anderson:1965}. Both algorithms must be \emph{vectorized}, i.e.~their
transformation must affect the entire solution at all momenta $k$ uniformly,
because the convergence speed would otherwise differ at different $k$ and 
the solution $M(k)$ would become progressively distorted by the acceleration.

\begin{figure}[t]
	\centering
	\includegraphics[width=0.44\linewidth]{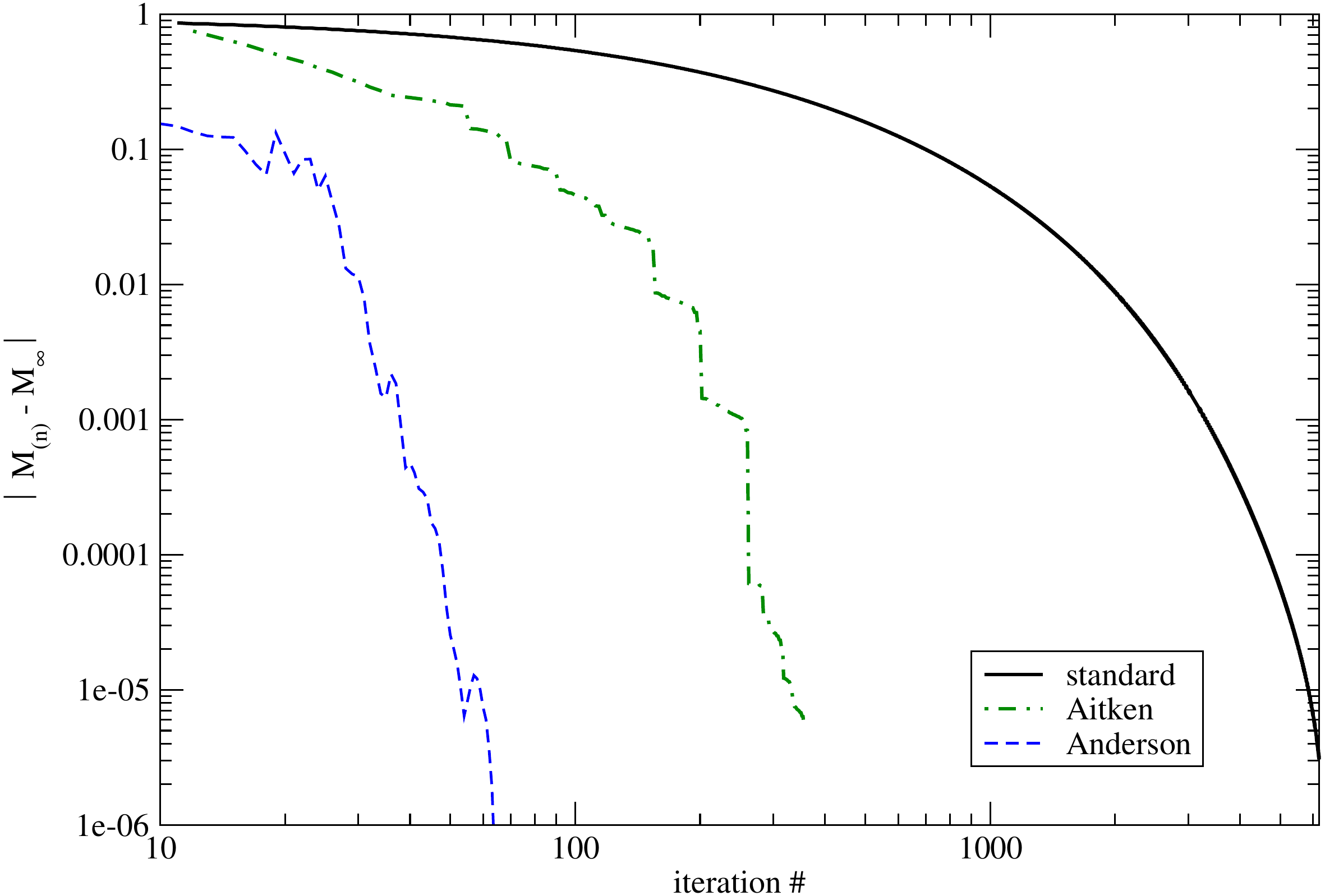}
	\hspace*{1cm}
	\includegraphics[width=0.42\linewidth]{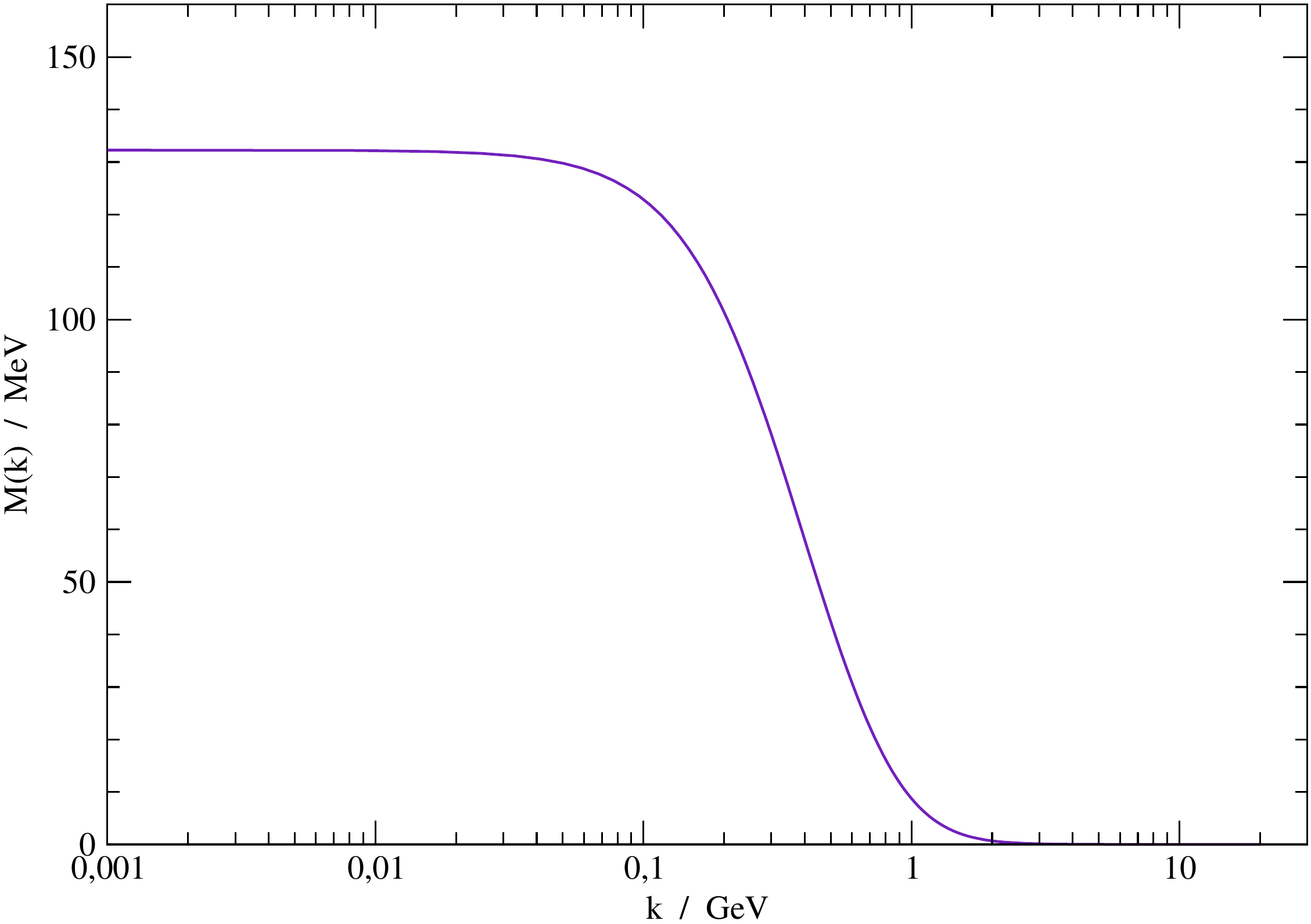}
	\caption{\emph{Left}: Convergence history of the standard and accelerated 
	iteration method. \emph{Right}: Solution of the gap equation for the mass 
	function $M(k)$ at $T=0$.}
	\label{fig:1}
\end{figure}

The accelerators generally buffer a certain number of iteration elements, 
and predict an improved estimator for the next iteration based on its history.
Anderson's method, in particular, comes with a level $k \ge 1$ that describes
the dimension of the sub-space in which the univariate secant method
is applied. It requires to store $(k+1)$ previous iterations, both 
accelerated and un-accelerated, and must solve a $k \times k$ linear system
for each iteration. Usually, $k=2$ and $k=3$ give the best results, while 
levels $k \ge 6$ rarely show any improvement. The Adler-Davis equation is 
different, however: Since it converges extremely slowly, the iteration can 
benefit from a much larger level $k$, combined with a moderate overrelaxation.
We found that $k=18$ with an overrelaxation $\alpha = 1.5$ give the best 
results. The outcome is pretty impressive: The unaccelerated iteration 
requires more than $7,000$ iterations to reduce the residual (the distance 
of the lhs and rhs of the integral equation)\footnote{Here and in the following, 
we are measuring the distance of solutions by the normalized $L_2$ metric, 
	\begin{align}
	\|M - X\|^2 \equiv \frac{1}{N} \sum_{i=0}^{N-1} \big[M(k_i) - X(k_i)\big]^2\,,
	\label{metric}
	\end{align}
where $\{ k_i \}$ are the grid positions on which the solutions are defined, 
and $N$ is the number of grid points. (This formula can be extended to
2D grids for mass functions $M(k,\xi_k)$ at finite temperature in an obvious 
manner.) The $L_2$ norm is a good compromise which measures convergence on
average while still giving each individual grid position enough weight so that 
outliers will not go by unnoticed.}
below a threhsold of $10^{-6}$, even when using overrelaxation. 
If we combine it with Aitken's method, the overall iteration count is reduced to 
about $400$ at the same accuracy, while Anderson's method requires only $66$ 
iterations to reach a residual of $10^{-9}$.  The sequence accelerator thus 
gives a higher final accuracy and easily saves us a factor $100$ of CPU time 
in the present case. We have plotted this situation again in the left panel of 
Fig.~\ref{fig:1},  where we show the iteration history, 
i.e.~the distance of the intermediate result at iteration \#$n$ to the final 
solution, $\| M_{(n)} - M_\infty \|$. As can be seen from the double logarithmic 
plot, the convergence of the accelerated sequences is less smooth but much faster 
than the standard iteration, and Anderson's method is clearly superior. 

The resulting solution to the $T=0$ equation is shown in the right panel of 
Fig.~\ref{fig:1}. As mentioned in the introduction, this mass function originates 
from the instantaneous part of the quark propagator in Coulomb gauge. It 
cannot be compared directly to the constituent mass function in Landau gauge,
and attempts to match the two definitions reveal that the infrared limit 
$M(0) \approx 133 \,\mathrm{MeV}$ observed here could still be compatible with the 
standard findings in Landau gauge \cite{Campagnari:2018flz}. The mass function 
computed here gives rise to a chiral condensate \cite{RV2016} 
\begin{align}
\langle \bar{\psi}\psi\rangle_0 = - \frac{N_\mathrm{C}}{\pi^2}\int\limits_0^\infty 
\dd p\,p^2\,\frac{M(p)}{\sqrt{p^2 + M(p)^2}}
\approx - (185\,\mathrm{MeV})^3\,,
\label{fatso}
\end{align}  
if the standard scale eq.~(\ref{scale}) corresponding to $\sigma_{\rm C} / \sigma = 2.5 $
is used.

\subsection{The finite temperature case: Poisson resummation}
At non-zero temperatures, the presence of the heat bath singles out 
a rest frame and the original spatial $O(3)$ symmetry is broken to $O(2)$. As explained 
earlier, we have put the heat bath in the spatial $3$-direction and compactified 
this dimension. The use of polar coordinates $\vk = (k_z, k_\perp, \varphi_k)$
is, however, discouraged since the mass function would then depend on two 
non-compact coordinates $k_z$ and $k_\perp$, which complicates the UV and, 
in particular, the IR limit considerably. A better strategy is to keep the 
spherical coordinates $(k, \vartheta_k, \varphi_k)$. The remaining axial 
$O(2)$ symmetry of rotations about the $3$-axis entails that we can place
the external momentum into the $1,3$-plane and set the azimuthal angle 
$\varphi_k = 0$. Also, the mass function must be invariant under the 
reflection $k_z \to (-k_z)$, as a remainder of the original $O(3)$ 
symmetry\footnote{This corresponds to the reflection $\nf \to -(\nf+1)$ of 
the Matsubara indices which flips the sign of the Matsubara frequency.} and we can take 
$\xi_k \equiv \cos \vartheta_k \ge 0$ without loss of generality. The mass 
function is hence
\begin{align}
M = M(k, \xi_k)\,,\qquad\qquad k \ge 0\,,\qquad\qquad \xi_k = \cos \vartheta_k 
\in [0,1]\,.
\end{align}
For the loop integration, we also adopt spherical coordinates
$(q, \vartheta, \varphi)$. The angles only enters through their cosine 
via the scalar product
\begin{align}
\vq \cdot \vk &= q_z\, k_z + \vq_\perp\cdot\vk_\perp 
= q_z\, k_z + q_\perp k_\perp\cos\varphi 
\nonumber \\[2mm]
&= q k \Big[\cos \vartheta\,\cos \vartheta_k + \cos\varphi\,\sqrt{1-\cos^2\vartheta}
\sqrt{1 - \cos^2 \vartheta_k}\Big] = qk \Big[ \xi \xi_k + \eta\Big]\,,
\end{align}
where we have introduced the cosines
\begin{align}
\xi_k = \cos \vartheta_k \in [0,1] \,,\qquad\qquad \xi = \cos \vartheta \in [-1,1] 
\,,\qquad\quad \gamma = \cos \varphi \in [-1,1] 
\end{align}
and defined the useful abbreviation
\begin{align}
\eta\equiv \gamma \sqrt{(1-\xi_k^2)(1-\xi^2)}\,.
\end{align}
In these coordinates, the Poisson re-summed gap equation (\ref{gap0}) in quotient form
becomes
\begin{align}
M(k,\xi_k) &= \frac{\displaystyle \frac{1}{\pi^2} \sum_{m=-\infty}^\infty
(-1)^m\, \int_{-1}^1 \dd\xi  \int_0^\infty \dd q \,
q^2 \,U(q)\,\cos\big[\beta m (k \xi_k + q \xi)\big]\,
\int_{-1}^1 \frac{\dd\gamma}{\sqrt{1-\gamma^2}}\,
\frac{M(Q,\xi_Q)}{\sqrt{Q^2 + M(Q,\xi_Q)^2}}}
{\displaystyle 1  + \frac{1}{\pi^2} \sum_{m=-\infty}^\infty
(-1)^m\,  \int_{-1}^1 \dd\xi \int_0^\infty \dd q \,
q^2\,U(q)\,\cos\big[\beta m (k \xi_k + q \xi)\big]\,
\int_{-1}^1 \frac{\dd\gamma}{\sqrt{1-\gamma^2}}\,
\frac{1 + q/k\,(\xi \xi_k + \eta)}{\sqrt{Q^2 + M(Q,\xi_Q)^2}}}
\label{gap1}
\end{align}
with the shifted momentum
\begin{align}
Q^2 = q^2 + k^2 + 2 k q\, (\xi\xi_k + \eta)\,, \quad\qquad\qquad
\xi_Q = |k \xi_k + q \xi| / Q\,.
\label{QQ}
\end{align}
If we compare this to the $T=0$ version in eq.~(\ref{zero}), it is evident that 
the $\mf=0$ term in the Poisson sums reproduces the $T\to 0$ limit, if we assume
that the mass function does not depend on $\xi_k$ because of the restored 
$O(3)$ symmetry. Also, the shifted momentum eq.~(\ref{QQ}) agrees with the $T=0$ limit 
in the equation below (\ref{zero}) when $\xi_k = 1$, i.e.~when the external momentum
points into the direction of the heat bath. This direction of the external momentum
therefore gives the  closest analogue of the $T=0$ mass function,
and we will therefore compare $M(k,\xi_k=1)$ to the $T=0$ limit in 
section \ref{sec:results} below. 

Eq.~(\ref{gap1}) is not yet suited for numerical investigation. As we have explained
in the previous section, a small regulator $\mu$ for the infrared divergence of the 
quotient form is required and provides for a stable iteration,
\begin{align}
U(q) = \frac{1}{(q^2 + \mu^2)^2}\,.
\label{regpot}
\end{align} 
At finite 
temperatures, however, the infrared divergence also leads to a poor behaviour of the Poisson 
series, whose terms typically decay very slowly when $\mu \ll 1$. It is therefore 
convenient to subtract an analytic helper function in the integrands of 
eq.~(\ref{gap1}), which will render the $q$-integral IR finite at $q \to 0$. 
Of course, we have to add back in what we subtracted, and this extra term will now 
carry the infrared divergence and the poor Poisson sum. The advantage of this procedure 
is that the part of the calculation which depends on the (numercially expensive) mass 
function is finite and quickly converging, while the problematic terms can be handled 
analytically. The gap equation now takes the form

\begin{align}
M(k,\xi_k) &= \frac{\displaystyle g(k,\xi_k) + \frac{1}{\pi^2} \sum_{m=-\infty}^\infty
	(-1)^m\, \int_{-1}^1 \dd\xi  \int_0^\infty \dd q \,
	q^2 \,U(q)\,\cos\big[\beta m (k \xi_k + q \xi)\big]\,
	\int_{-1}^1 \frac{\dd\gamma}{\sqrt{1-\gamma^2}}\,u(q,\xi,\gamma\,;\,k,\xi_k)}
{\displaystyle h(k,\xi_k)  + \frac{1}{\pi^2} \sum_{m=-\infty}^\infty
	(-1)^m\,  \int_{-1}^1 \dd\xi \int_0^\infty \dd q \,
	q^2\,U(q)\,\cos\big[\beta m (k \xi_k + q \xi)\big]\,
	\int_{-1}^1 \frac{\dd\gamma}{\sqrt{1-\gamma^2}}\,v(q,\xi,\gamma\,;\,k,\xi_k)}\,,
\label{gapx}
\end{align}
where the integrands read
\begin{align}
u(q,\xi,\gamma\,;\,k,\xi_k) &= \frac{M(Q,\xi_Q)}{\sqrt{Q^2 + M(Q,\xi_Q)^2}} 
- \Delta u(q,\xi,\gamma\,;\,k,\xi_k)  
\nonumber \\[2mm]
v(q,\xi,\gamma\,;\,k,\xi_k) &= \frac{1 + q/k\,(\xi \xi_k + \eta)}{\sqrt{Q^2 + M(Q,\xi_Q)^2}} - \Delta v(q,\xi,\gamma\,;\,k,\xi_k)\,,
\label{integrands}
\end{align}
and the subtractions are compensated by the inhomogeneities $g$ and $h$. An obvious 
choice for the subtractions is the $q=0$ limit of the integrands,
\begin{align}
\Delta u(q,\xi,\gamma\,;\,k,\xi_k) &= \frac{M(k,\xi_k)}{\sqrt{k^2 + M(k,\xi_k)^2}}  
\nonumber \\[2mm]
\Delta v(q,\xi,\gamma\,;\,k,\xi_k) &= \frac{1}{\sqrt{k^2 + M(k,\xi_k)^2}}\,.
\label{subtract}
\end{align}
With this subtraction, the integrands $u$ and $v$ behave as $\mathcal{O}(q)$ at 
small momenta, but the leading $\mathcal{O}(q)$ term is linear in $\gamma$ and 
thus $\gamma$-integrates to zero. The result of the $\gamma$-integration 
is hence $\mathcal{O}(q^2)$ which, together with the factor $q^2 U(q) \sim q^{-2}$, 
yields a finite loop integral at $q \to 0$, even in the absence of a cutoff.
The infrared divergence now re-appears in the inhomogeneities
\begin{align}
g(k,\xi_k) &= \frac{1}{\pi^2} \sum_{\mf = -\infty}^{\infty} (-1)^\mf \int\limits_{-1}^1 \dd\xi
\int\limits_0^\infty \dd q\,q^2 U(q) \cos(\beta \mf(k \xi_k + q \xi)) \int\limits_{-1}^1 \frac{\dd\gamma}
{\sqrt{1-\gamma^2}} \,\Delta u(q,\xi,\gamma\,;\,k,\xi_k)
\nonumber\\[2mm]
h(k,\xi_k) &= 1 + \frac{1}{\pi^2} \sum_{\mf = -\infty}^{\infty} (-1)^\mf \int\limits_{-1}^1 \dd\xi
\int\limits_0^\infty \dd q\,q^2 U(q) \cos(\beta \mf(k \xi_k + q \xi)) \int\limits_{-1}^1 \frac{\dd\gamma}
{\sqrt{1-\gamma^2}} \,\Delta v(q,\xi,\gamma\,;\,k,\xi_k)\,,
\end{align}
where it can be handled analytically: after performing the integrations with a small
infrared regulator $\mu > 0$ in the potential $U(q)$ as in eq.~(\ref{regpot}), we obtain
\begin{align}
g(k,\xi_k) &= \frac{M(k,\xi_k)}{\sqrt{k^2 + M(k,\xi_k)^2}} \sum_{\mf=-\infty}^\infty
(-1)^\mf\,\frac{\cos(\beta \mf k \xi_k)}{2\mu}\,\exp(-\mf \beta \mu)
\nonumber \\[2mm]
&=\frac{M(k,\xi_k)}{\sqrt{k^2 + M(k,\xi_k)^2}} \cdot \frac{1}{2\mu}\,
\frac{\sinh(\beta \mu)}{\cosh(\beta \mu) + \cos(\beta k \xi_k)}\,.
\label{inhomo}
\end{align}
The calculation for $h$ is identical, without the overall factor $M(k,\xi_k)$ 
in the numerator,
\begin{align}
h(k,\xi_k) &= 1 + \frac{1}{\sqrt{k^2 + M(k,\xi_k)^2}} \sum_{\mf=-\infty}^\infty
(-1)^\mf\,\frac{\cos(\beta \mf k \xi_k)}{2\mu}\,\exp(-\mf \beta \mu)
\nonumber \\[2mm]
&=1 + \frac{1}{\sqrt{k^2 + M(k,\xi_k)^2}} \cdot \frac{1}{2\mu}\,
\frac{\sinh(\beta \mu)}{\cosh(\beta \mu) + \cos(\beta k \xi_k)}\,.
\label{inhomo1}
\end{align}
Note that the full Poisson sum appears to be \emph{finite} when the regulator 
is removed, while the $\mf=0$ term is $1/(2\mu)$ and thus really diverges in the infrared.
As explained in the previous section, both numerator and denominator of the gap 
equation in quotient form \emph{should} be infrared divergent at $T=0$ and this 
property should also persist at $T>0$. This apparent contradiction comes from an order 
of limits issue:
since the relevant terms in eq.~(\ref{inhomo}) have the argument $\beta \mu$, the limit 
$\mu \to 0$ at finite $\beta$ gives a finite result, while the limit $\beta \to \infty$ at 
finite $\mu$ yields the expected $T=0$ divergence $1 / (2\mu)$. This indicates that 
the correct formulation (the one which is continuously connected to the $T=0$ case) must 
retain a small, but finite IR cutoff $\mu > 0$. Taking $\mu \to 0$ too early will lead 
to a different (finite) formulation in which the $T \to 0$ limit disagrees with the 
well-known $T=0$ result. We will thus always keep a small but non-zero IR cutoff 
$\mu > 0$ which ensures a smooth limit $T\to 0$ and also stabilizes the iteration 
as explained in the previous section.

With the Fourier integrand going as $\mathcal{O}(q^0)$ at $q \to \infty$, simple dimensional
analysis suggests that the terms in the Poisson sum decay as $(\beta m)^{-1}$ which,
together with the alternating sign, amounts to a poorly converging series.\footnote{Even with a 
regulator, the damping factor $e^{- \beta |\mf| \mu}$ indicates that of the order 
$1 / (\beta \mu)\approx 10^5$ terms would have to be summed if done naively.} 
This type of slowly converging alternating series can, however, be handled quite 
efficiently using the $\epsilon$-algorithm \cite{Wynn:1952,Wynn:1956}. The alternative 
would be to attempt one more subtraction of the $\mathcal{O}(q^2)$ behaviour under the 
integrands $u$ and $v$. This leads, however, to a rather formidable expression involving
up to second order derivatives of the mass function. Since the latter is only known 
numerically on a rather coarse momentum grid, the second subtraction 
cannot be carried out with sufficient accuracy and we stick to eq.~(\ref{subtract}).

Eq.~(\ref{gapx}) is the final form of the gap equation 
which we solve iteratively: we start with an arbitrary function $M_0(k)$, either the 
$T=0$ solution or a constant, 
\begin{align}
M(k,\xi_k) = M_0(k)\qquad\qquad \forall\quad \xi_k \in [0,1]\,,
\label{init}
\end{align}
and use Anderson's algorithm as a sequence accelerator as in the T=0 case, cf.~fig.~\ref{fig:1}. 
Once the system has been iterated to convergence, we can extract the chiral condensate from 
\begin{align}
\langle \bar{\psi}\psi\rangle &= - \frac{\Ncc}{\pi^2} \sum_{\mf = -\infty}^\infty 
(-1)^\mf \int\limits_0^1 \dd\xi \int\limits_0^\infty \dd q\,q^2\,\cos(\mf \beta q\,\xi)
\frac{M(q,\xi)}{\sqrt{q^2 +  M(q,\xi)^2}}
\label{Xcond}
\end{align}
which is the finite temperature extension of eq.~(\ref{fatso}).

\subsection{The finite temperature case: Matsubara formulation}
\label{matsu}
The poisson resummation technique described in the last section is convenient 
at low temperatures were only a few terms are required. In addition, the $T=0$ limit 
is recovered from the lowest term of the series. As the temperature increases, more
and more terms of the Poisson series have to be included. At very high temperatures, 
we may thus reach a point were the original Matusbara formulation is more convenient,
as the relevant sums as saturated by the first few Matsubara frequencies. In particular, 
only the lowest Matsubara frequency survives in the high-temperature limit $T\to \infty$. 

Though our numerical procedure mainly relies on the Poisson technique, we have also 
solved quark gap equation (\ref{gapeq}) in the Matsubara representation and compared 
it with the results of the Poisson formulation. This will provide an independent test 
for the accuracy of our numerics. 

For the Matsubara formulation, we employ the residual $O(2)$ symmetry of rotations 
about the $3$-axis to let the external momentum component in the plane perpendicular
to the heat bath point into 1-direction, $\mathbf{k}_\perp = k_\perp\,\mathbf{e}_1$.
For the loop integration, we use polar coordinates $q_\perp$ and 
$\xi = \cos\sphericalangle(\mathbf{q}_\perp,\mathbf{e}_1)$. We can then express 
eq.~(\ref{gapeq}) in these coordinates, scale all dimensionfull quantities in the 
units of eq.~(\ref{scale}) and finally go over to the more stable quotient form. 
This gives
\begin{align}
M(k_\perp, \Omega_\ell) = \frac{\displaystyle \frac{2}{\pi \beta} 
\sum_{\nf=0}^\infty \int_0^\infty \dd q_\perp \int_{-1}^1 \dd\xi\,
\frac{q_\perp}{\sqrt{1-\xi^2}}\sum_{\pm} U\Big(\sqrt{q_\perp^2 + (\Omega_\ell \pm \Omega_\nf)}\Big)\,
\frac{M(Q_\perp, \Omega_\nf)}{\sqrt{Q_\perp^2 + \Omega_\nf^2 + M(Q_\perp, \Omega_\nf)^2}}}
{\displaystyle 1 +  \frac{2}{\pi \beta} 
	\sum_{\nf=0}^\infty \int_0^\infty \dd q_\perp \int_{-1}^1 \dd\xi\,
	\frac{q_\perp}{\sqrt{1-\xi^2}}\sum_{\pm} U\Big(\sqrt{q_\perp^2 + (\Omega_\ell \pm \Omega_\nf)}
	\Big)\,
	\frac{\textstyle  1 + \frac{k_\perp q_\perp \xi +
			\Omega_\ell(\pm \Omega_n - \Omega_\ell)}{k_\perp^2 + \Omega_\ell^2}}{\sqrt{Q_\perp^2 + \Omega_\nf^2 + M(Q_\perp, \Omega_\nf)^2}}}
\label{gapmats}
\end{align} 
where $Q_\perp \equiv \sqrt{k_\perp^2 + q_\perp^2 + 2 k_\perp q_\perp \xi}$. 
Because of the symmetry $M(k_\perp, \Omega_\ell) = M(k_\perp, -\Omega_\ell) = 
M(k_\perp, \Omega_{-\ell-1})$, we can restrict the external Matsubara index to 
$\ell \ge 0$. On the rhs of eq.~(\ref{gapmats}), we have also combined the terms 
with index $\nf$ and $-(\nf+1)$, since they only differ in the sign of the 
frequency $\Omega_\nf$. This ensures that only mass functions with Matsubara index 
$\nf \ge 0$ appear on both sides of eq.~(\ref{gapmats}) and the coupled integral 
equation system closes. The chiral condensate can be expressed in the Matsubara 
formulation as 
\begin{align}
\langle \bar{\psi} \psi \rangle = - \frac{2 \Ncc}{\pi \beta}\sum_{\nf=0}^\infty 
\int\limits_0^\infty \dd q_\perp \,q_\perp\,\frac{M(q_\perp, \Omega_\nf)}{\sqrt{q_\perp^2 + 
\Omega_\nf^2 + M(q_\perp, \Omega_\nf)^2}}\,.
\label{matscond}
\end{align}

For a numerical evaluation, the potential $U(q)$ must be infrared regularized as in
eq.~(\ref{regpot}), and the number of Matsubara frequencies included in the system 
must be restricted to $\nf < N$. The system (\ref{gapmats}) then resembles the $T=0$ 
equation, however with an $N$-component solution $M_\ell(k_\perp) \equiv M(k_\perp, \Omega_\ell)$
and a different integration measure. It is the latter property which makes the Matsubara 
formulation less convenient: for small regulators $\mu \ll 1$, the potential $U$ has a 
strong singularity at $q_\perp = 0$ if the external and loop frequency match, 
$\Omega_\ell = \Omega_n$. This singularity is only partially cancelled by the integration 
measure and the $\nf=\ell$ term  dominates the entire Matsubara sum by a 
relative factor $1/\mu^2$. As before, this singular factor is canceled between the 
numerator and denominator,
and the remaining $\mathcal{O}(\mu^2)$ contributions from the other terms $\nf \neq \ell$ 
in the Matsubara sum carry the actual corrections to the mass functions. This means 
that the iteration progresses much slower than at $T=0$ and, more problematic, the 
$\nf \neq \ell$ terms from the Matsubara sums must be computed to a very high 
accuracy. 
 
In addition to the high accuracy demand, the Matsubara formulation has the 
property that all components $M(k_\perp, \Omega_\ell)$ are coupled by the system 
(\ref{gapmats}), so that the index cutoff $N$ must be fixed once and for all 
and cannot be adjusted dynamically.\footnote{It is possible to increase the number 
of frequencies included in the sum by assuming that $O(3)$ invariance is restored 
for large frequencies. This allows to approximate $M(k_\perp, \Omega_\ell) \approx
M(\bar{k}_\perp, \Omega_{N-1})$ for large frequencies $\ell \ge N$, where 
$\bar{k}_\perp = \sqrt{k_\perp^2 + \Omega_\ell^2 - \Omega_{N-1}^2} > k_\perp$.
For small $N$, this approximation is not applicable, and for large $N$ it is 
unnecessary since the sum will converge without it. The extrapolation is 
hence most useful in the intermediate region $N\approx 40$ where it can save a 
factor of 2-4 in computation time.}
If the external index $\ell$ approaches the cutoff $N$, there are only a few 
frequencies larger than the dominating contribution $\nf = \ell$, i.e. 
the Matsubara sum is truncated unsymmetrically and the higher frequencies 
$\ell \approx N$ thus have a systematic bias. The only solution is to 
include a very large number of frequencies $N$, so that the inaccurate 
modes near the cutoff give such a small contribution that their combined 
error does not matter. Unfortunately, the computational effort of the system
(\ref{gapmats}) scales strictly as $\mathcal{O}(N^2)$, so that the inclusion of 
higher frequencies is limited by practical considerations. In our studies, we 
were able to push the Matsubara mode count to $N=100$ for the lower temperatures,
which means that we have $10,000$ times the numerical effort of the $T=0$ solution 
per iteration, plus a much larger iteration count due to the slow convergence of 
eq.~(\ref{gapmats}) and an increased accuracy demand. Even with this considerable effort,
the frequency count was just enough to map the transition region, but temperatures 
below $T = 50\,\mathrm{MeV}$ give incorrect results and require a different (massively 
parallel) strategy, cf.~Fig.~\ref{fig:8}. 
By contrast, the Poisson formulation -- though much more 
costly per iteration -- scales better with increasing temperature as it can adjust 
its mode count dynamcially and hence will \emph{always} give a correct result, 
albeit with a (moderately) increased effort at higher temperatures. 

Finally, it should also be mentioned that the convergence of the Matsubara system 
(\ref{gapmats}) is non-uniform in the frequency index, i.e.~the lowest frequencies 
are stable after a relatively small iteration count, while the highest frequencies
(which contribute the least) are the slowest to converge. In practice, we have to 
stop the iteration at some point where the highest frequencies may not have fully 
converged. Since we relax from above, this means that the highest frequency mass 
functions are systematically too large, and, although each frequency contributes 
very little, their combined effect may lead to overestimate the condensate 
eq.~(\ref{matscond}).
We will see this effect in Fig.~\ref{fig:8} below, where the Matsubara values 
for the condensate in the transition region are all slightly larger than the 
corresponding Poisson results.

\section{Results}
\label{sec:results}

We split our result section in two parts: first we consider numerical details 
on the individual parts of our calculation to demonstrate that the fairly 
complicated process actually works as intended. In the second part, we discuss
the final results for the mass function and the chiral condensate at 
different temperatures. 

\begin{figure}[t!]
	\centering
	\includegraphics[width=0.4\linewidth]{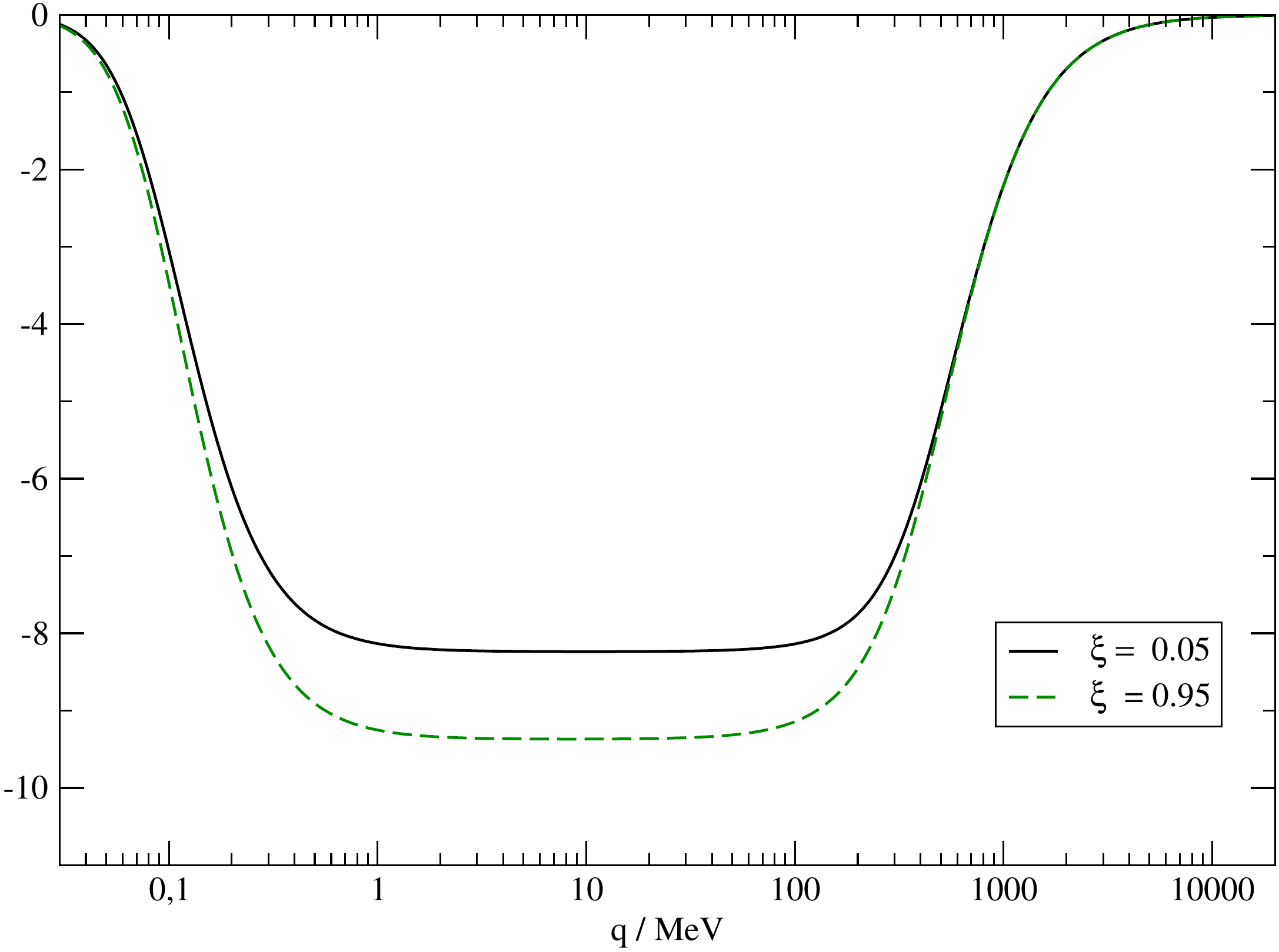}
	\qquad\qquad
	\includegraphics[width=0.385\linewidth]{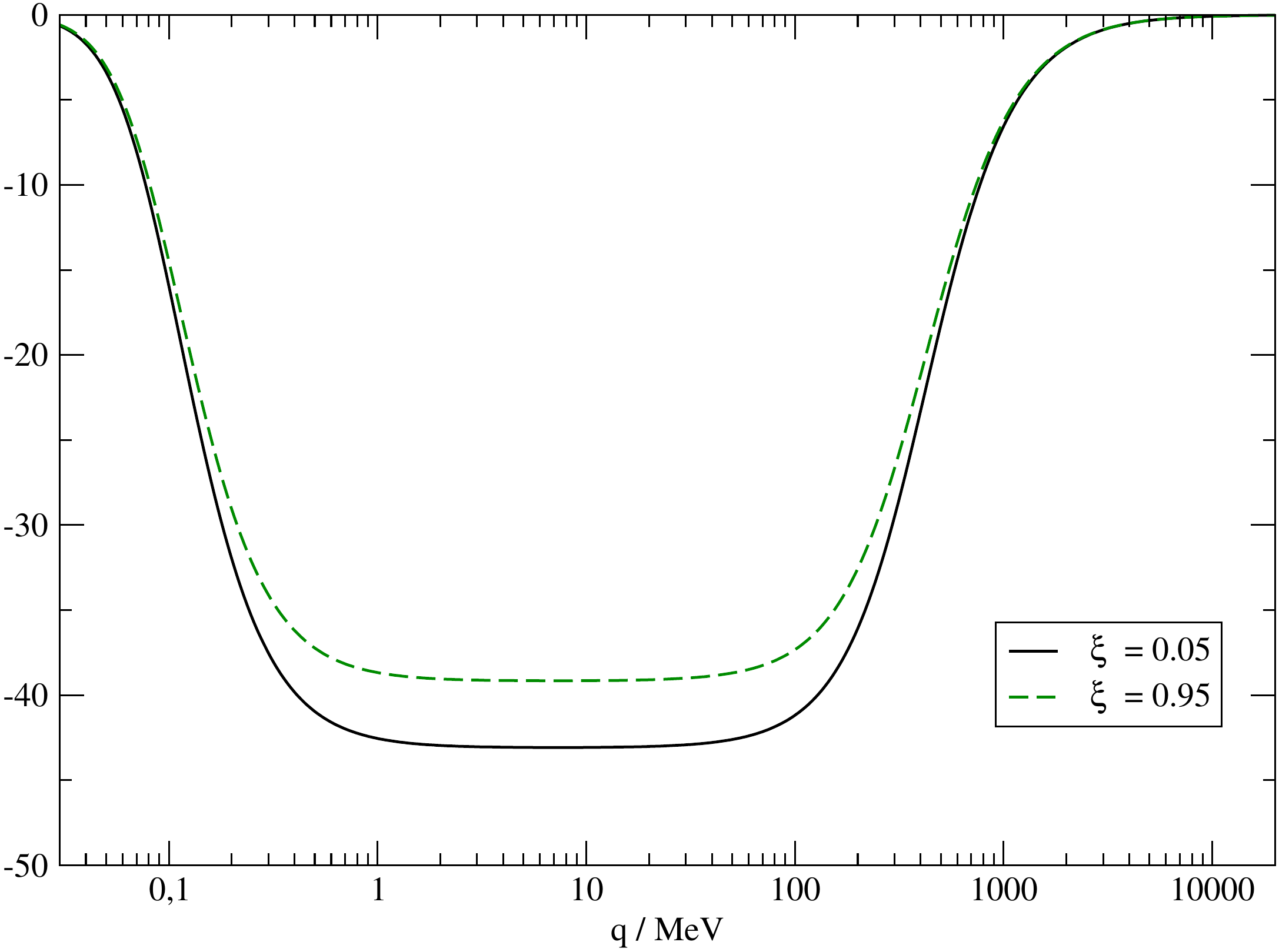}	
	\caption{Integrand of the momentum integral of eq.~(\ref{gapx}) without the Fourier cosine 
		factor, cf.~eq.~(\ref{pic3}). The two panels show the numerator 
		(\emph{left}) and denominator (\emph{right}) of the gap equation (\ref{gapx}), 
		respectively. Parameters are $\xi = 0.05$ and $\xi=0.95$ for the 
		loop angle, and $k = 200\,\mathrm{MeV}$ and $\xi_k = 0.5$ for the 
		external momentum.}
	\label{fig:3}
\end{figure}

\subsection{Details on the numerical method}
\label{subsec:details}
In the following, we present some typical results of intermediate steps in the 
calculation. Numerical issues appear predominantly in the earlier steps of the 
iteration and the eventual mass function has a similar shape to the initial 
zero-temperature solution, cf. below. For simplicity, we can therefore assume 
the zero-temperature mass function for the qualitative arguments in this 
subsection. Furthermore, we fix the external momentum to a typical value 
$k = 200\,\mathrm{MeV}$ and $\xi_k = 0.5$, which is in the region where the 
mass function changes most quickly.

It should also be noted that we generally combine terms with both signs 
$\pm \xi$ in all internal calculations, i.e.~we symmetrize the 
$\xi$-integrand
\begin{align}
\int\limits_{-1}^1 \dd\xi\, f(\xi) = \int\limits_0^1  \dd\xi \,\big[ f(\xi) + f(-\xi)\big]\,.
\label{symmetrize}
\end{align}
To keep the formulas simple, we will not always indicate this symmetrization,
which is implicitly understood.

\begin{figure}[t!]
	\centering
	\includegraphics[width=0.4\linewidth]{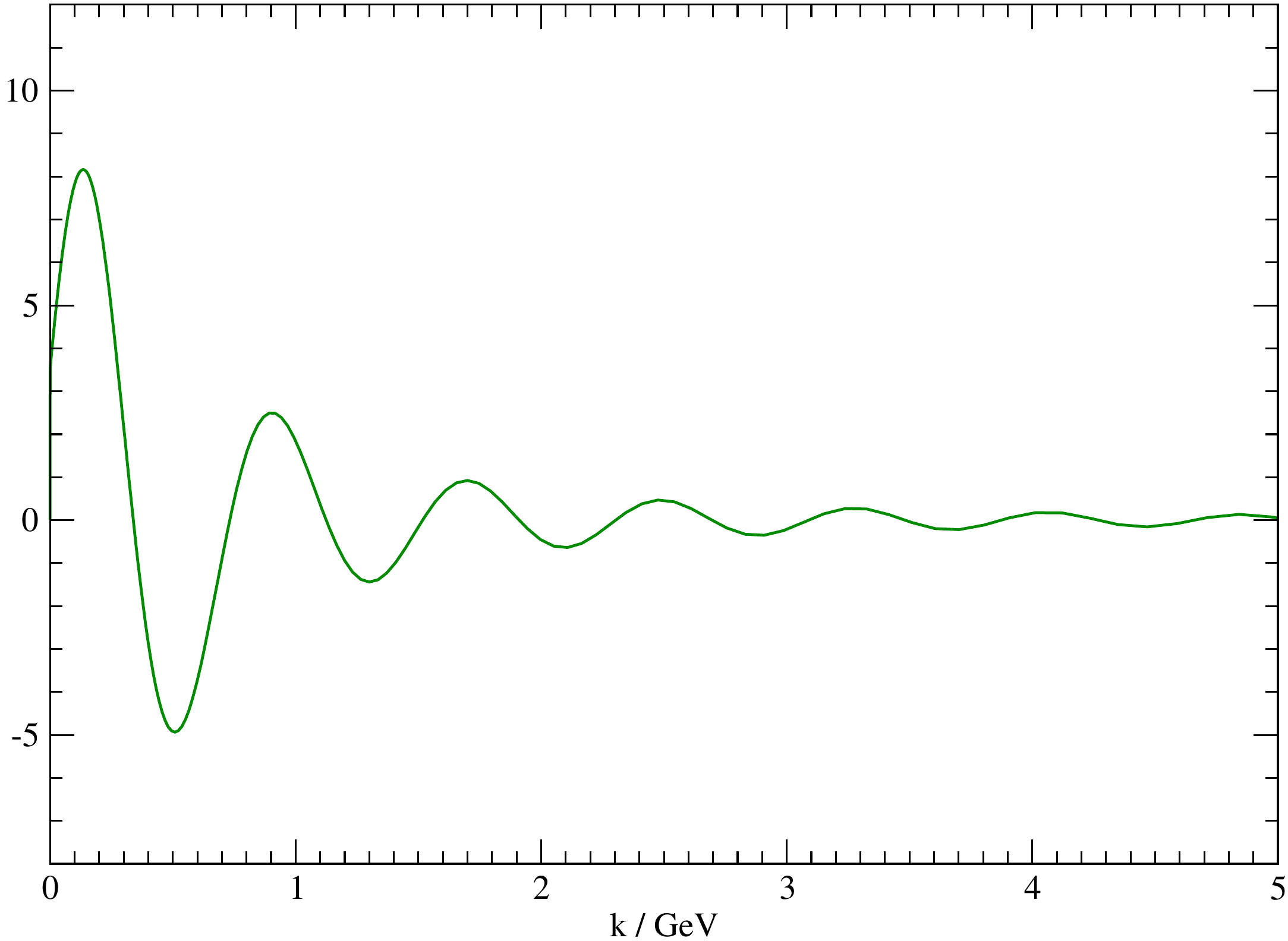}
	\qquad\qquad
	\includegraphics[width=0.4\linewidth]{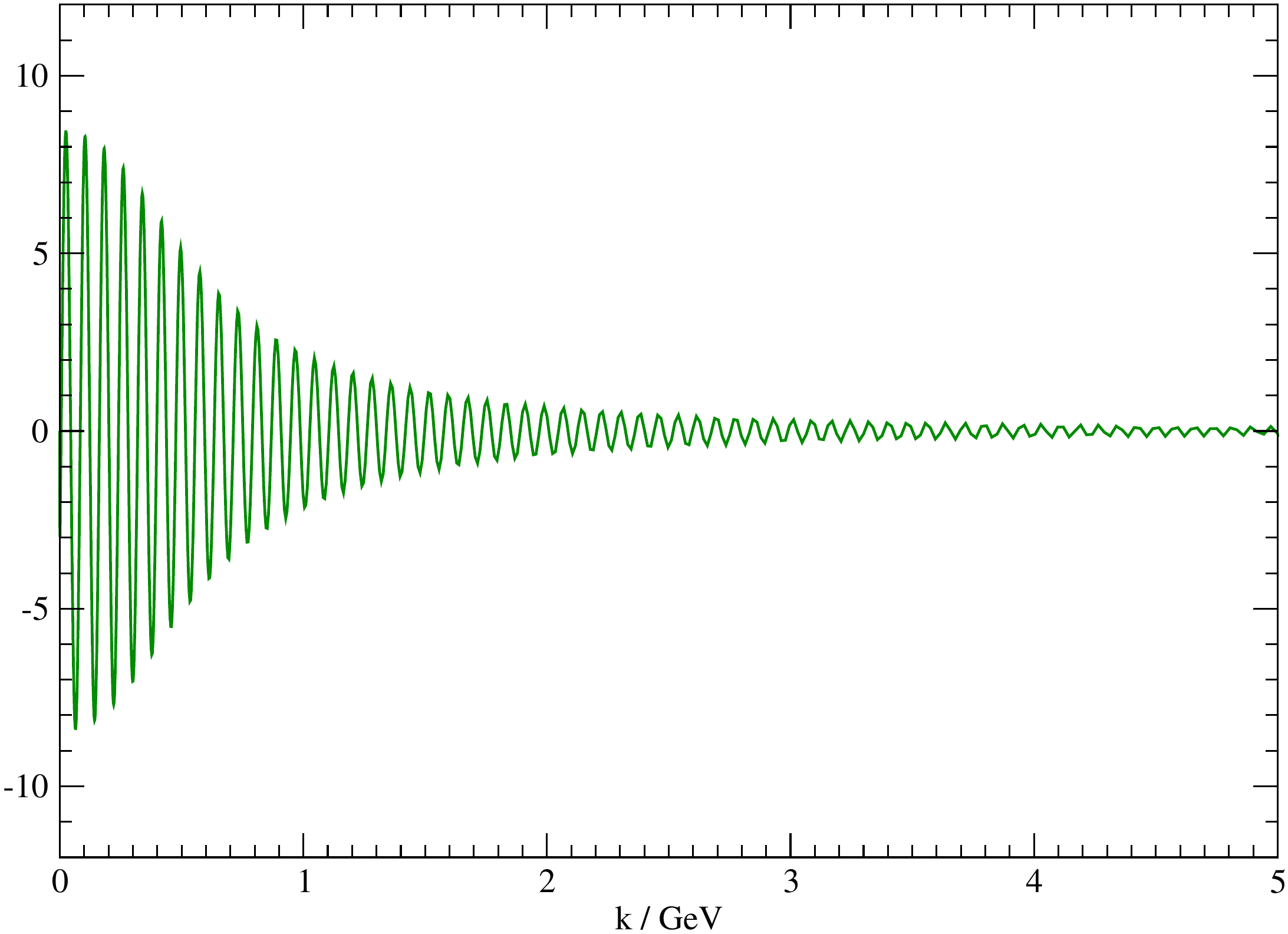}	
	\caption{Full integrand of the momentum integral in the numerator
		of the gap equation (\ref{gapx}) for a temperature of 
		$T = 50 \,\mathrm{MeV}$ and Poisson index $\mf = 1$ (\emph{left}) 
		and $\mf = 10$ (\emph{right}).}
	\label{fig:4}
\end{figure}

We begin with the integrand of the momentum integral omitting the Fourier 
cosine factor for clarity and combining terms from $\xi$ and $(-\xi)$,
\begin{align}
q \mapsto q^2 U(q)\int\limits_{-1}^1 \frac{\dd\gamma}{\sqrt{1-\gamma^2}} 
\sum_{\pm} u(q,\pm \xi,\gamma\,;\,k,\xi_k)\,.
\label{pic3}
\end{align}
The integration here can be done very efficiently using 
\emph{Gauss-Chebychev} integration, which automatically takes care of 
the square root factor in the denominator.
In Fig.~\ref{fig:3}, we have plotted eq.~(\ref{pic3}) for two values 
of $\xi$ close to the boundary, for both the numerator (left panel) and 
the denominator (right panel) of the gap equation  (\ref{gapx}).
The shape of the functions is quite similar in all cases: for small 
momenta $q$, the functions approach a constant, which is maintained for 
about two to three orders of magnitude, before the influence of the 
regulator $\mu = 0.1 \,\mathrm{MeV}$ sets in and the functions quickly 
vanish for $q < \mu$. For the numerator, the functions with 
the larger $\xi$ lies above the one with the smaller $\xi$, while this 
order if reversed in the denominator. The general shape of all these 
functions is in agreement with our discussion of the subtraction above. 
The actual integrand of the momentum integral still has the Fourier 
cosine factor, which leads to the oscillating functions depicted 
in Fig.~\ref{fig:4}.

\begin{figure}[t!]
	\centering
	\includegraphics[width=0.4\linewidth]{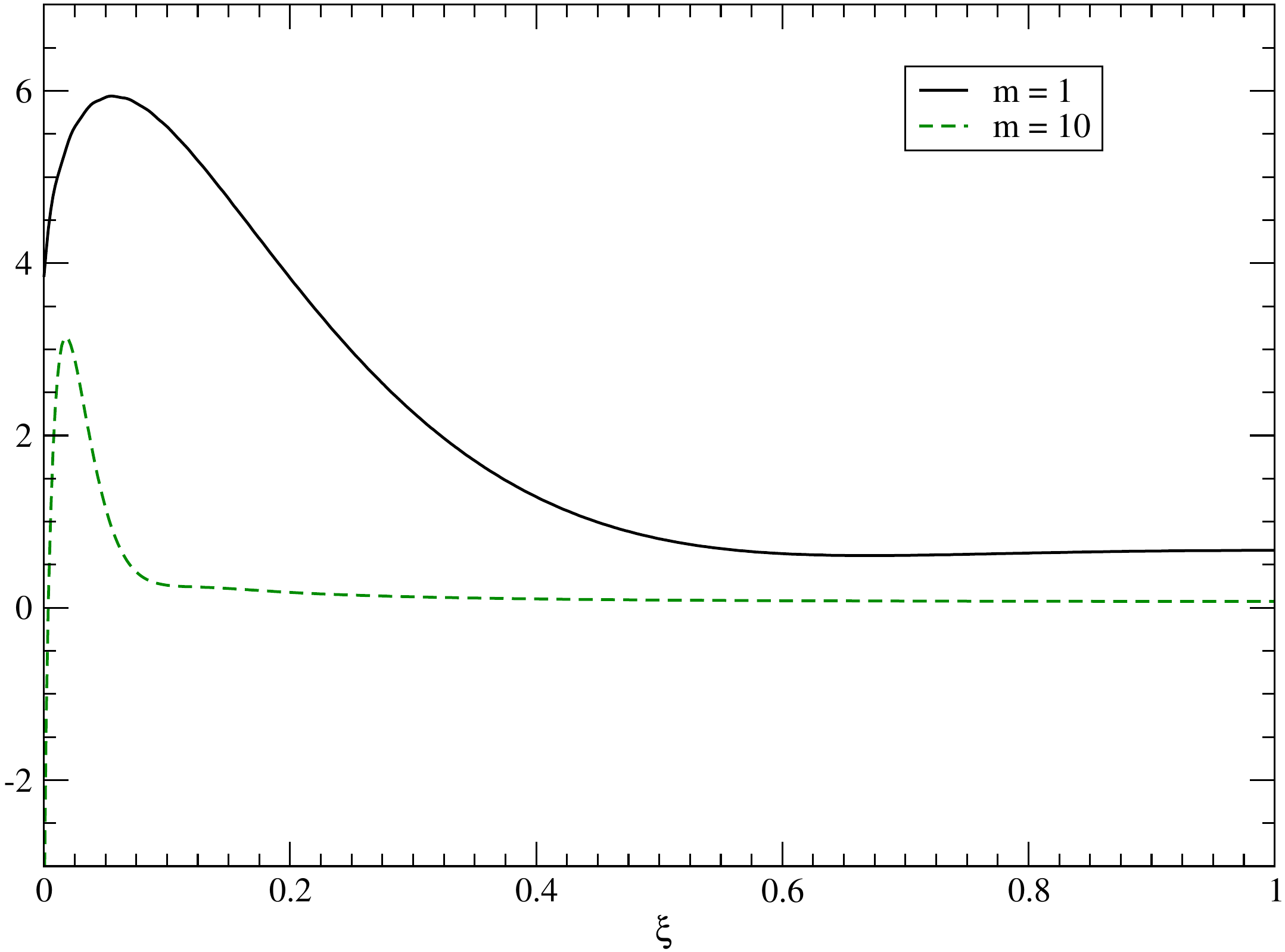}
	\qquad\qquad
	\includegraphics[width=0.4\linewidth]{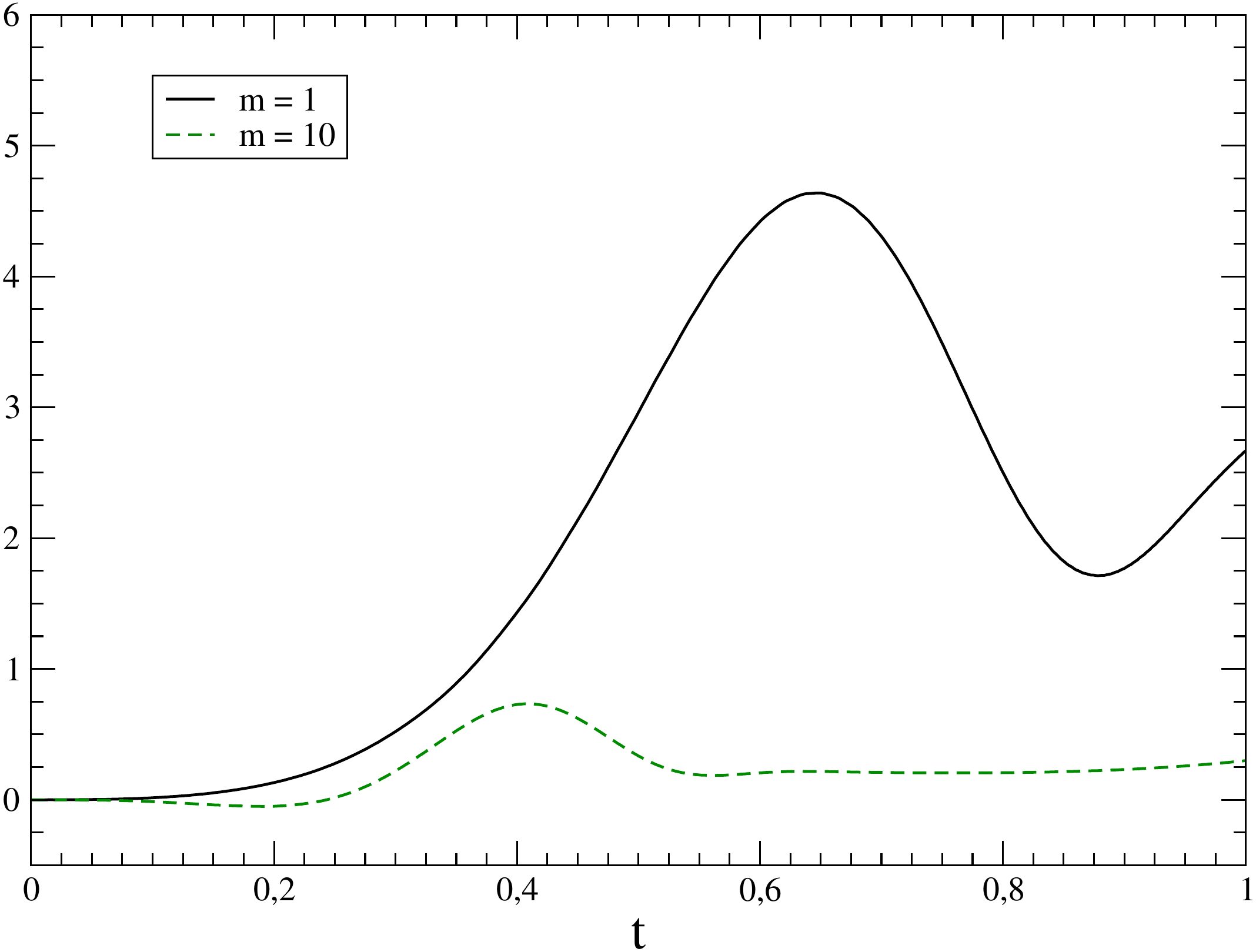}	
	\caption{\emph{Left:} Integrand of the $\xi$-integral in the numerator of 
		the gap equation gap equation (\ref{gapx},) for a temperature of 
		$T = 50 \,\mathrm{MeV}$ and two Poisson indices $\mf = 1$ and $\mf=10$. 
		\emph{Right:} The same functions after applying the change of 
		variables $t = \xi^4$.}
	\label{fig:5}
\end{figure}

\medskip
\noindent
Next, we consider the integrand of the $\xi$-integral after performing the 
Fourier momentum integration,
\begin{align}
\xi \mapsto \int\limits_0^\infty \dd q\,q^2\, U(q)\,\cos(\beta \mf (k \xi_k + q \xi))
\,\int\limits_{-1}^1 \frac{\dd\gamma}{\sqrt{1-\gamma^2}}\,\sum_{\pm} 
u(q, \pm \xi, \gamma\,;\,k, \xi_k)\,.
\label{xintegrand}
\end{align}
Note that the argument $\xi$ in this function can be restricted to $\xi \in [0,1]$
due to the symmetrization of $\pm \xi$. Besides the external momentum (which we have 
fixed to the same standard value as in the previous figures), this function now 
depends on both the temperature and the Poisson summation index. In the left 
panel of Fig.~\ref{fig:5}, we have plotted eq.~(\ref{xintegrand}) for a fixed 
temperature $T=50\,\mathrm{MeV}$ and two Poisson indices $\mf=1$ and $\mf=10$. 
As can be seen, the function eq.~(\ref{xintegrand}) is non-oscillating and 
rather smooth, except for a steep drop in the vicinity of $\xi=0$. This is the 
region where the Fourier momentum integral has a low frequency and thus the 
cancellations due to rapid oscillations are absent. Numerically, an accurate 
integration of eq.~(\ref{xintegrand}) requires a large number of sampling points
if a uniform $\xi$-sampling is adopted. Alternatively, it is more efficient to 
spread out the low $\xi$-behaviour by a change of variables $\xi = t^n$ with 
$n > 1$,
\begin{align}
\int\limits_0^1 \dd\xi \,f(\xi) = \int\limits_0^1 \dd t\, n t^{n-1} f(t^n) \equiv 
\int\limits_0^1 \dd t\,\bar{f}(t)\,.
\label{tfromxi}
\end{align}

\noindent
In the right panel of Fig.~\ref{fig:5}, we show the $n=4$
transformation eq.~(\ref{tfromxi}) of the plots in the left panel. The detailed 
structure at low $\xi$ is now spread out and the resulting function can be 
accurately integrated using Gauss-Legendre with about $30$ sampling points.
(By contrast, more than $1200$ uniformly distributed sampling points were used
for the left panel of Fig.~\ref{fig:5}). The absolute value of the $\xi$- 
integrals in the left panel of Fig.~\ref{fig:5} are $1.933$ for $\mf=1$ and 
$0.221$ for $\mf=10$, respectively. This demonstrates the relatively slow 
$1/\mf$ decay of the Poisson sum, even at a small temperature of 
$T=50\,\mathrm{MeV}$. The integral of the transformed function in the 
right panel of Fig.~\ref{fig:5} agree, of course, with the corresponding 
integral in the left panel. The integrands in the denominator of the gap equation
show a qualitatively similar behavior and are not plotted here for brevity.

\begin{figure}[t!]
	\centering
	\includegraphics[width=0.4\linewidth]{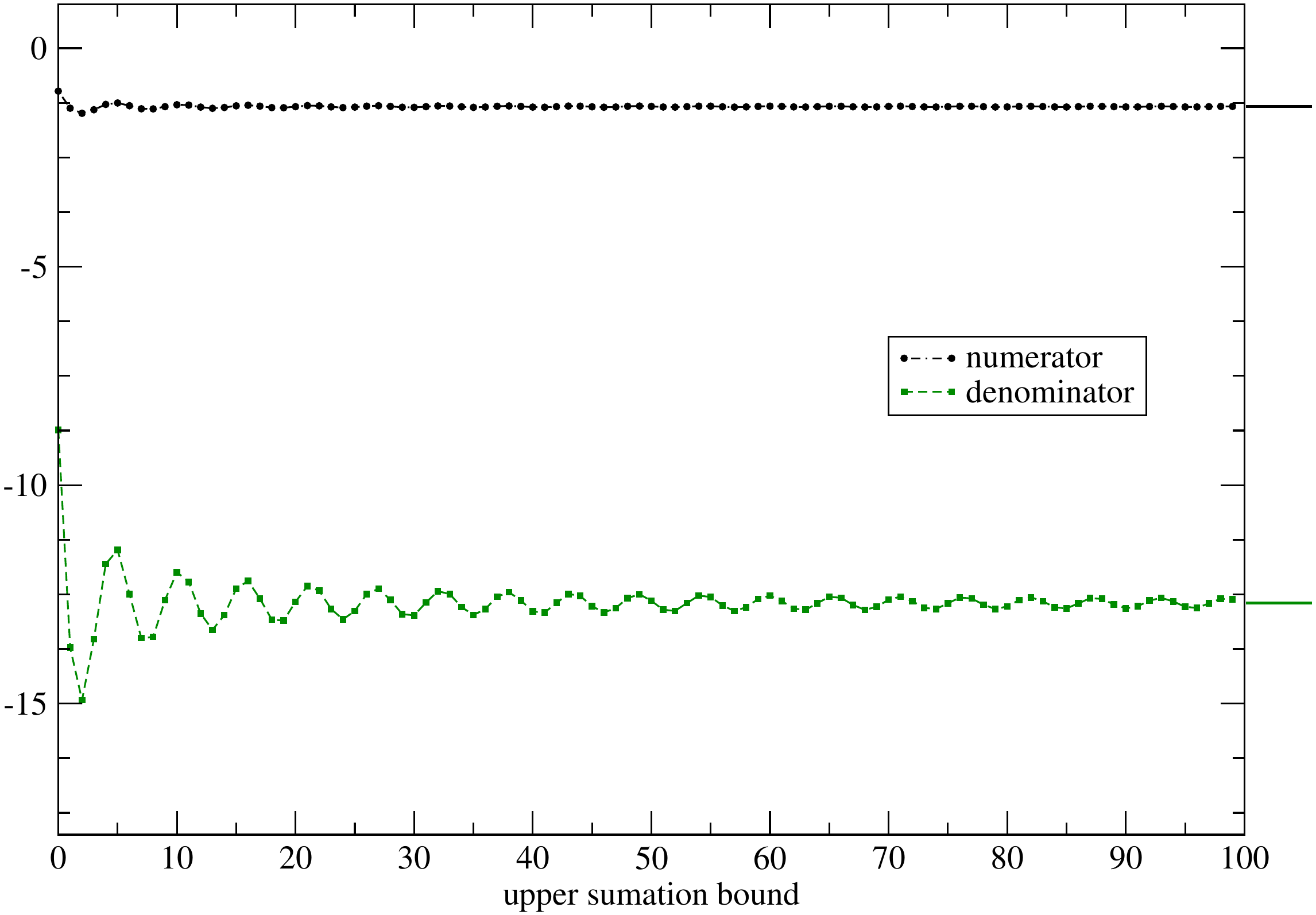}
	\qquad\qquad
	\includegraphics[width=0.4\linewidth]{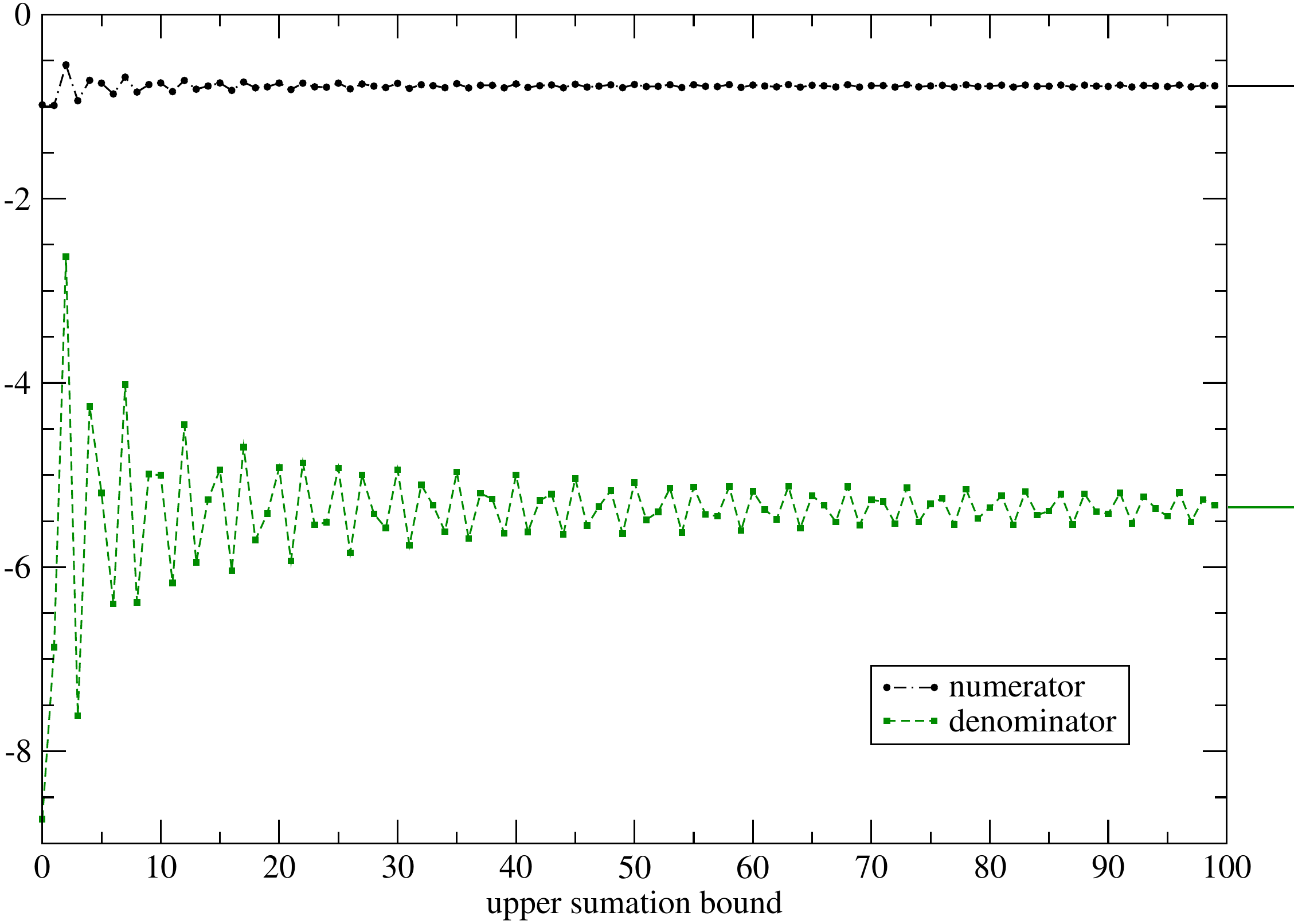}	
	\caption{Partial sums in the numerator and denominator of the gap equation
	(\ref{gapx}), as a function of the upper summation bound. The small horizontal 
	bar on the right of the coordinate box indicates the value for the infinite series
	predicted by the $\epsilon$-algorithm. The left panel is for $T=50\,\mathrm{MeV}$,
	while the right panel shows $T=150\,\mathrm{MeV}$. In all cases, the external momentum 
	was fixed to the preferred value $k=200\,\mathrm{MeV}$ and $\xi_k = 0.5$.}
	\label{fig:6}
\end{figure}

\medskip\noindent
Finally, we check the convergence of the Poisson sums in the gap equation
(\ref{gapx}). We fix the external momentum again at our preferred value 
$k = 200\,\mathrm{MeV}$ and $\xi_k = 0.5$, and plot the partial Poisson 
sums in eq.~(\ref{gapx}), including the prefactor $1/\pi^2$. 
The sums are even in $\mf$, i.e. we can combine terms with $\pm \mf$, 
\beq
\sum_{\mf= -\infty}^\infty a_\mf = \sum_{\mf = 0}^\infty (2 - \delta_{\mf 0})\,
a_\mf \equiv \sum_{\mf=0}^\infty b_\mf\,.
\eeq
Fig.~\ref{fig:6} presents the partial sums in the gap equation (\ref{gapx}) as a 
function of the upper summation bound, for our preferred external momentum setup.
The left panel shows the situation for $T=50\,\mathrm{MeV}$, while the right 
panel displays $T=150\,\mathrm{MeV}$. The $\mf=0$ term dominates in all cases,
while the $\mf > 0$ terms contribute with alternating signs, giving rise to
oscillations in the partial sum which decay rather slowly ($\sim \mf^{-1}$). 
These oscillations are much more pronounced for the higher temperature in the 
right panel of Fig.~\ref{fig:6}. Numerically, we have found that up to 
$10,000$ terms would have to be summed in this case, in order to suppress 
the oscillations and predict the value of the infinite series to a relative 
accuracy of $10^{-4}$. By contrast, the $\epsilon$-algorithm is able to reach 
the same accuracy from only the first $25$ terms in the series. 

\subsection{Results}
\label{subsec:results}

% As explained earlier, the closest analog of the $T=0$ mass function $M_0(k)$ in
% our finite temperature formulation is $M(k, \xi_k=1)$, when the external momentum
% points in the direction of the heat bath. This special configuration has the 
% highest symmetry under the residual axial $O(2)$ rotations around the heat 
% bath, and the finite temperature gap equation (\ref{gapx}) reduces to the $T=0$ 
% equation if $\xi_k=1$ and only the $\mf=0$ part of the Poisson series is taken.

\begin{figure}[t!]
	\centering
	\includegraphics[width=0.45\linewidth]{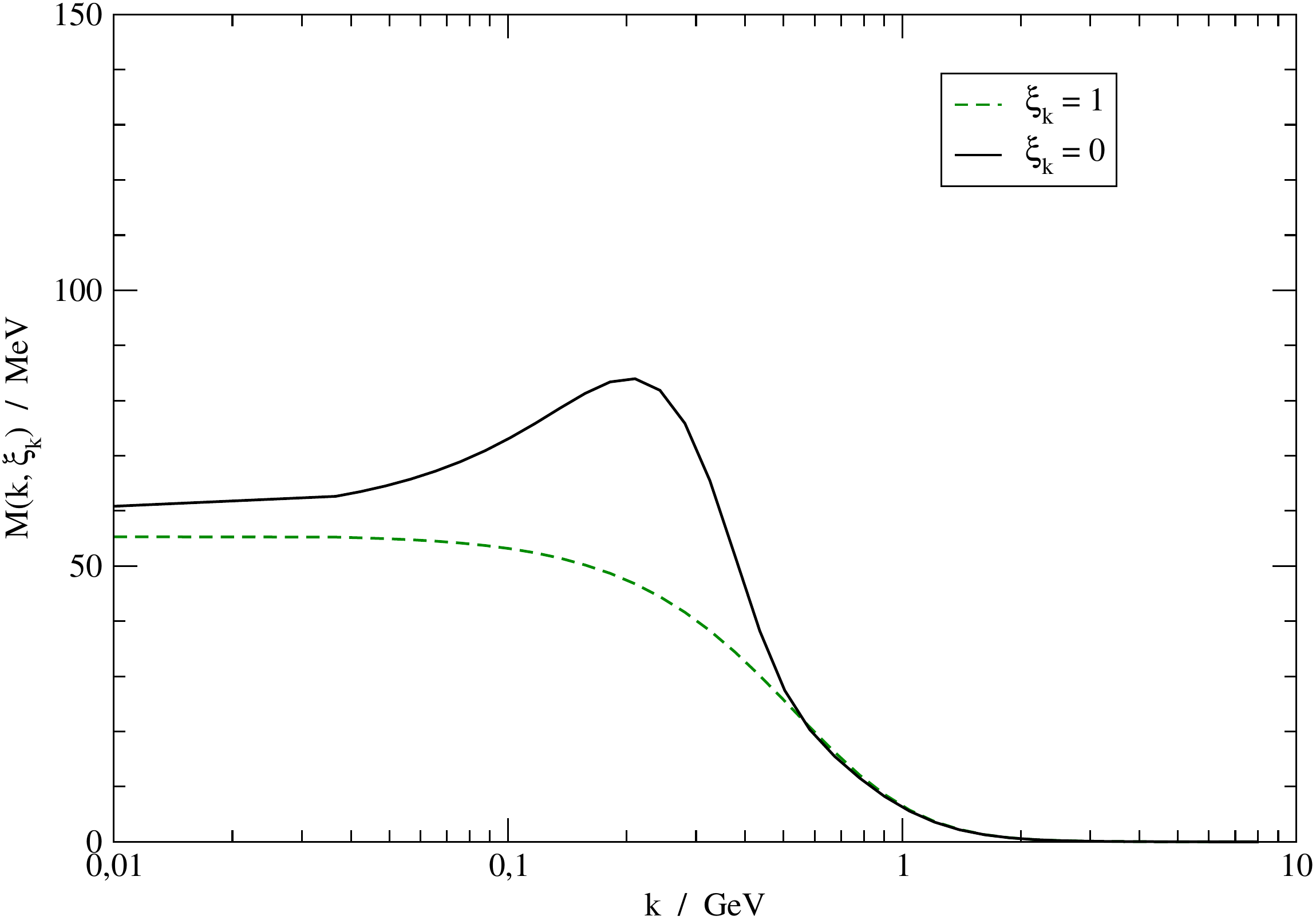}
	\qquad\qquad
	\includegraphics[width=0.45\linewidth]{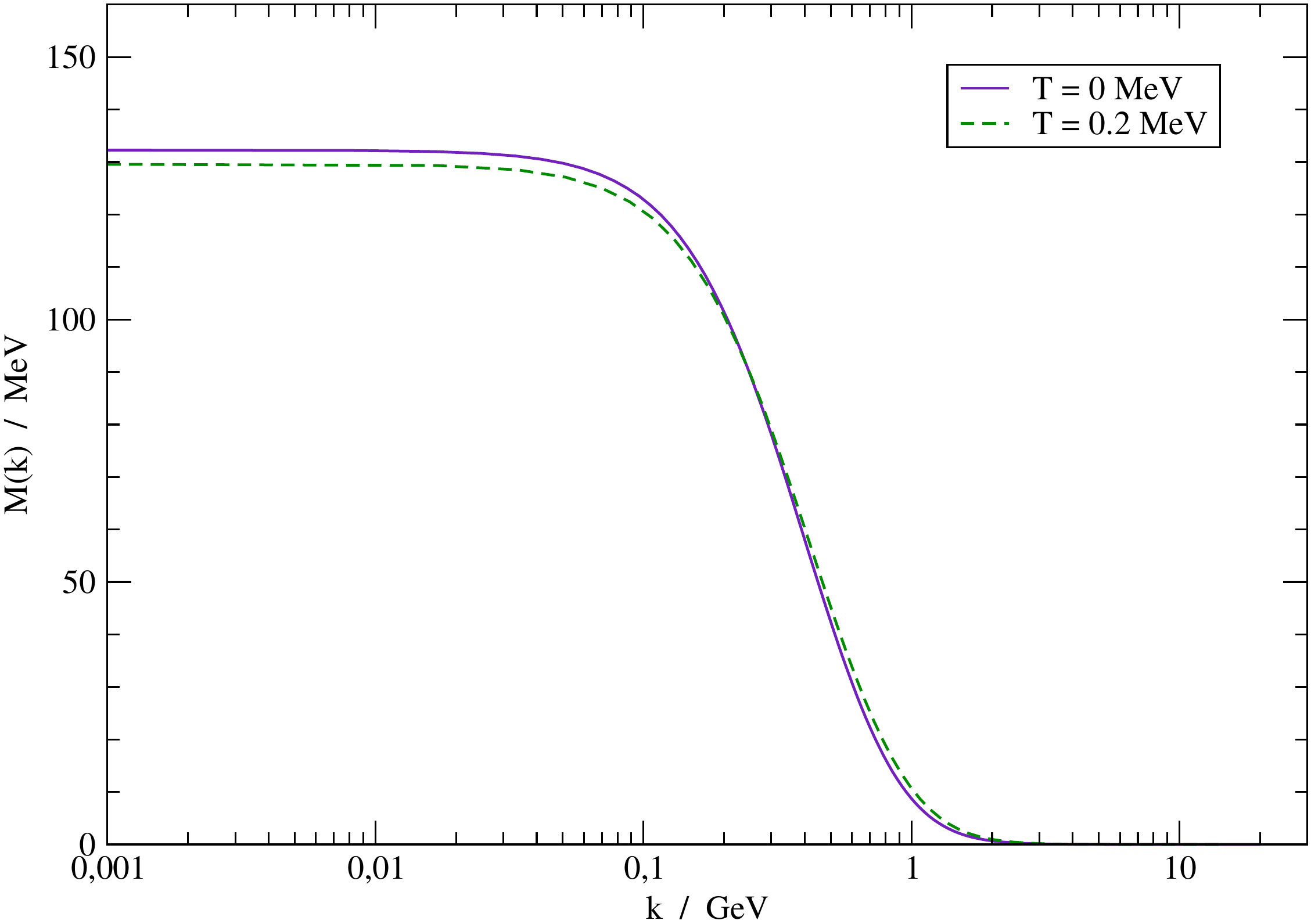}	
	\caption{\emph{Left:} Mass function $M(k,\xi_k)$ at $T=80\,\mathrm{MeV}$ 
	with the momentum $\vk$ pointing in various directions relative to the heat bath.
	\emph{Right:} Mass function $M(k,1)$ for small temperatures compared to the 
	$T=0$ limit.}
	\label{fig:7}
\end{figure}

In the left panel of Fig.~\ref{fig:7}, we show the mass function $M(k,\xi_k)$ for the 
two extreme directions $\xi_k$ of the external momentum (longitudinal and perpendicular 
to the heat bath). As explained earlier, the closest analog of the $T=0$ mass function
$M_0(k)$ in our finite temperature formulation is $M(k, \xi_k=1)$, when the external momentum
points in the direction of the heat bath. This curve has indeed a very similar form to the 
$T=0$ solution plotted for comparison, while the $\xi=0$ curve shows some deviations. 
Also, the mass function is already considerably smaller than at $T=0$, even though the 
temperature $T=80\,\mathrm{MeV}$ in this plot is still in the confined phase. In order to 
recover the $T\to0$ limit, we would thus have to go to very small temperatures. This is 
shown in the right panel of Fig.~\ref{fig:7}, where we compare the mass function 
for small temperatures with the $T=0$ limit.
% \footnote{The violation of $O(3)$ symmetry 
% for these small temperatures is very small, and $M(k,\xi_k=0)$ would essentially give the 
% same curve.} 
As can be seen, the mass function starts to drop already at temperatures below 
$10\,\mathrm{MeV}$. From eq.~(\ref{Xcond}), this does not directly translate into a drop 
of the condensate, since the mass function appears in the numerator and denominator 
of the integrand, which is thus less sensitive to the mass function at small momenta. 
(The temperature also appears through the Fourier sum which gives the main temperature 
sensitivity.) 

For these reasons, the intercept $M(0,1)$ of the mass function is not a good 
indicator for the phase transition, in particular since it is also gauge dependent.
This is also the reason why it cannot be directly compared to the constitutent mass 
in conventional covariant gauges; instead, it represents a gauge-dependent 
\emph{mass parameter} from the instantaneous part of the quark propagator 
in Coulomb gauge. In Ref.~\cite{Campagnari:2018flz} it was demonstrated that such a
mass parameter in Coulomb gauge could be as low as $M(0) \approx 100\,\mathrm{MeV}$ 
and still be compatible with the (much larger) constituent masses in Landau gauge. 

\begin{figure}[t!]
	\centering
	\includegraphics[width=0.8\linewidth]{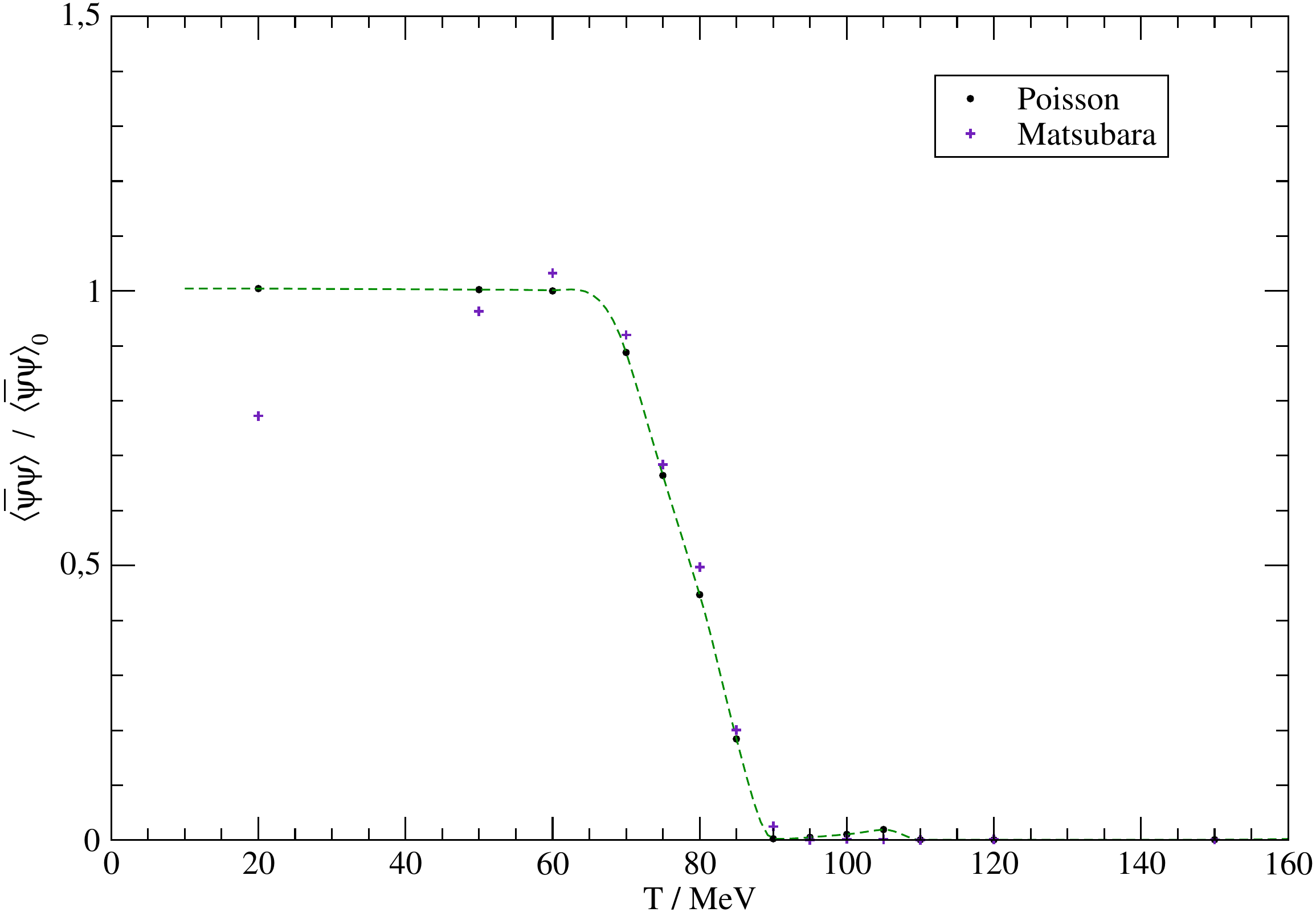}	
	\caption{Chiral condensate as a function of the temperature, from both 
	the Matsubara and Poisson formulation. The dashed line indicates a fit 
    to the Poisson data from which the critical temperature is determined.}
	\label{fig:8}
\end{figure}

The gauge-invariant order parameter for the phase transition is the chiral condensate
plotted in Fig.~\ref{fig:8}. This shows the expected behaviour, i.e. it is roughly 
constant for small temperatures, and drops quickly to very small values at a
characteristic transition temperature $T^\ast$. For higher temperatures, it is 
practically zero within our numerical precision.\footnote{For temperatures $T > T^\ast$,
the non-trivial solution presumably ceases to exist, and we only have the trivial
solution $M=0$, which is always present. The fact that the condensate is not exactly 
zero is a numerical artifact: the iterative solution requires very many iterations to 
relax to $M=0$, and we have simply stopped the process after about $100+$ CPU hours.}
The exact location of the phase transition temperature $T^\ast$ depends on 
the order of the transition: If it is a strong cross-over, one usually defines 
$T^\ast$ from the inflexion point, while a second order transition would have 
a jump discontinuity in the derivative at the temperature where a 
non-vanishing condensate appears for the first time as we cool from the  
deconfined phase. This value of the transition temperature is always larger 
than the inflexion point. Our data indeed indicates at a (weak) second order 
transition with a critical temperature of 
\begin{align}
T^\ast = 92\,\mathrm{MeV}\,.
\end{align} 
Both the order and the critical temperature differ from the result of our previous
investigation in ref.~\cite{RV2016}, where a broad crossover phase transition with
pseudo-critical temperature $165 \, \mathrm{MeV}$ was obtained in the limit of a 
vanishing quark-gluon coupling. It should, however, be mentioned that the solution 
of the zero temperature gap equation was used at all temperatures in Ref.~\cite{RV2016},
which is certainly a quite crude approximation.

In Fig.~\ref{fig:8}, we have also included data from a direct summation of the 
original Matsubara series (\ref{gapeq}) with the same cutoff $\mu = 0.1\,\mathrm{MeV}$
to take care of the infrared-singularity and the transformation to quotient form 
for better stability. This system scales quadratically with the number of included 
Matsubara frequencies, and up to $100$ frequencies (i.e. 10,000 times the effort 
as compared to $T=0$) were necessary to even reach the transition region.\footnote{In 
principle, an infinite number of frequencies would be necessary to overcome the infrared
singularity in the Matsubara formulation, and including only a finite number 
results in a regulator dependency that becomes most pronounced in the deep infrared, 
where the condensate for the Matsubara calculation clearly falls short of the expected 
value.}
By contrast, the Poisson formulation, though much more expensive per iteration, has no 
infrared singularity and provides the correct results even at larger temperatures,
where the increase in computational effort as compared to lower temperatures is still 
moderate. Combined with the physical transparency of the method, this warrants the 
high numerical effort of the Poisson resummation.
 
Returning to the results for the chiral phase transition, it should be emphasized again
that the absolute value of $T^\ast$ as well as 
the size of the condensate depends on the overall scale set by the Coulomb string 
tension, cf.~eq.~(\ref{scale}). The cited values are for the preferred value 
$\sigma_{\rm C} = 2.5\,\sigma$. In view of the uncertainties about the fundamental 
scale, it is better to cite our findings for the critical temperature as 
\begin{alignat}{3}
T^\ast / \sqrt{\sigma_{\rm C}}  = 0.13  \,.
\label{Tstar}
\end{alignat} 
For instance, a somewhat larger value $ \sigma_{\rm C} / \sigma = 4.1 $ would 
reproduce the lattice results for the chiral condensate and push the transition 
temperature to around $T^\ast \approx 118\,\mathrm{MeV}$.

A second-order chiral phase transition is the expected result for a system of \textit{two}
chiral quark flavors, as can be seen from the so-called Columbia diagram \cite{Karsch2002,
Fukushima2011}. This might be surprising since we consider only one single quark flavor within
our model. One should, however, notice that as long as the variational kernel $S$ is
flavor-diagonal, our results would not change even if we would take more flavors into account.
Since the neglect of the bare quark masses is the best approximation for two quark flavors
(i.e.~up and down), a second-order phase transition is definitely meaningful for the
considered model. Especially, it also agrees with the result found when finite temperatures
are introduced within the grand canonical ensemble, see e.g.~refs.~\cite{Alkofer1989, LS2010}.
The introduction of a flavor-dependent variational kernel might become necessary as soon as
finite current quark masses and unquenching effects are incorporated into the model. 
This could yield a dependence of the order of the phase transition on the number of quark 
flavors, as one would also expect from the Columbia diagram.

While the order of the chiral phase transition is the same in both the canonical and 
the present approach to finite temperatures, the critical temperature 
$T^{\ast}_{\mathrm{can}} \approx 64 \,\mathrm{MeV}$ found\footnote{Note that we adjusted the 
result from ref.~\cite{LS2010} to the value of the Coulomb string tension used in the 
present paper.} in the numerical calculations of ref.~\cite{LS2010} for vanishing chemical 
potential $\mu = 0$ is significantly smaller than our result. There are several possible 
reasons for this discrepancy: On the one hand, the evaluation of the partition function 
[see eq.~(\ref{Gl: DichteOperator})]
\beq
\mathcal{Z} = \tr \varrho = \tr \, \exp\bigl(-\beta [\mathcal{H} - \mu N]\bigr) \, ,
\eeq
where $\mathcal{H}$ is the QCD Hamiltonian on $\mathbb{R}^3$, necessitates some approximations
in the canonical approach. This concerns especially the treatment of the density operator
$\varrho$ [Eq.~(\ref{Gl: DichteOperator})] of the grand canonical ensemble, where a
quasi-particle approximation is required for the Hamiltonian $\mathcal{H}$.\footnote{Such an
approximation can be done by performing a Bogoliubov transformation of $\mathcal{H}$ and
keeping only the diagonal single-particle contributions \cite{Adler1984,
Davis1984}.}\textsuperscript{,}\footnote{The approach of ref.~\cite{Alkofer1989} corresponds
-- in the standard Hamiltonian approach -- to a quasi-particle ansatz for the density operator
of the grand canonical ensemble and minimizing the grand canonical potential with respect to
the quasi-particle energies.} Note that such an approximation is not required in the present
approach and no equations of motion for the quasi-particle energies hence emerge.
On the other hand, however, the present approach is based on the finite temperature formalism of 
Ref.~\cite{Reinhardt2016}, which in turn relies on the $O(4)$-invariance of the Lagrangian -- 
this is not entirely fulfilled in the Adler--Davis model since it contains the fermionic parts
of the Coulomb interaction (\ref{Gl: Coulombterm}) -- which is related to the $A_0$-$A_0$ 
correlator \cite{Greensite2003} -- but no contribution from the spatial gluons.

Nevertheless, the value for the critical temperature eq.~(\ref{Tstar}) is too small as
compared to the one found in lattice simulations for physical quark masses,
$T^{\ast}_{\mathrm{lat}} \approx 155 \, \mathrm{MeV}$ \cite{Borsanyi2010, *Bazavov2012}. 
Note that also the gauge invariant chiral quark condensate (Fig.~\ref{fig:8}) 
falls behind the value expected from phenomenology. The fact that the Adler--Davis model
predicts significantly too small results, e.g.~for the pion decay constant, is well-known
and also obtained in the canonical approach \cite{Alkofer2005}. In this respect, we should
stress that an increase of the critical temperature, and other quantities as the condensate,
is expected when the coupling of the quarks to the transverse spatial gluons is included, as
the numerical calculation performed in Ref.~\cite{RV2016} shows. Such a more complete system
would also allow for a more reliable comparison between the results obtained within the novel
and the canonical approach to finite-temperature Hamiltonian QCD.

Unfortunately, the inclusion of the coupling to the transversal spatial gluons will
drastically increase the numerical costs, which is already considerable in the present study. 
Although the coupling effects turned out to be small when the solution of the zero-temperature 
gap equation is used \cite{RV2016}, it is not clear if this is still true in the full 
temperature-dependent calculations. In contrast, the inclusion of finite bare quark masses should 
only smear out the phase transition to a crossover without having major effects on the value 
of the critical temperature. 

Finally, one should emphasize again that the Coulomb string tension could also be adjusted to 
$ \sigma_C  / \sigma \approx 4.1$, which reproduces the phenomenological value of the quark
condensate, $\langle \bar{\psi}\psi\rangle = - \big(236\,\mathrm{MeV}\big)^3$, but is somewhat 
larger than the lattice prediction. With this arrangement, one would find a critical temperature 
of $T^\ast = 118 \,\mathrm{MeV}$.

\section{Summary and Conclusions}
\label{sec:conclusions}
In the present paper, we have revisited the alternative Hamiltonian approach to
finite-temperature QCD of Ref.~\cite{Reinhardt2016} and solved the temperature-dependent 
equations of motion of the fermion sector numerically. In a first study, we have 
ignored the coupling of quarks and transverse spatial gluons. The apparent infrared
singularity of the resulting model could be resolved using Poisson resummation of the 
original Matsubara series, and the resulting integral equation system was solved 
using refined numerical techniques combined with analytical computations. Even 
though the iteration is stable and amenable to series accelerations, the loss of 
$O(3)$ symmetry requires a Poisson sum and three nested integrations per temperature, 
of which the Fourier quadrature, in particular, is fairly expensive. As a consequence, 
the final solution amounts to 100+ CPU hours per temperature.
   
For zero bare quark masses, the results for the chiral quark condensate show the expected 
weak second-order phase transition with a critical temperature of 
% TODO
$T^\ast\approx 92 \, \mathrm{MeV}$. 
While this value is larger than the one found within the usual (canonical) approach to 
finite temperature Hamiltonian QCD \cite{LS2010}, our findings for the critical 
temperature are definitely smaller than the result of lattice calculations using 
dynamical quarks,  $T^{\ast}_{\mathrm{lat}} \approx 155 \, \mathrm{MeV}$
\cite{Borsanyi2010, *Bazavov2012}.

We suspect that the mismatch between our findings and the lattice results is related 
to the neglect of the quark-gluon coupling in the variational ansatz, which also leads 
to a value of the quark condensate which is too small. However, if the scale is adjusted 
to reproduce the physical value of the quark condensate instead of fixing the scale 
from the rather poorly determined Coulomb string tension,
the critical temperature increases to about $118\,\mathrm{MeV}$, which is much
closer to the lattice data (while $\sigma_C / \sigma$ is still in the range supported 
by lattice calculations.) In forthcoming investigations, we therefore 
intend to study how this coupling affects the chiral phase transition. The solution of 
the full coupled equations of motion will also allow for the evaluation of the 
dressed Polyakov loop as order parameter for confinement. 
Furthermore, we plan to extend our calculations to the general case of a finite 
chemical potential, $\mu \neq 0$, in order to obtain a description of the whole 
QCD phase diagram within the Hamiltonian approach.

\section*{Acknowledgments}

This work was partially supported by 
Deutsche Forschungsgemeinschaft (DFG) under Contract No.~DFG-Re856/9-2 and under contract 
DFG-Re856/10-1.
 
\bibliography{thermalAD}

\begin{thebibliography}{44}%
\makeatletter
\providecommand \@ifxundefined [1]{%
 \@ifx{#1\undefined}
}%
\providecommand \@ifnum [1]{%
 \ifnum #1\expandafter \@firstoftwo
 \else \expandafter \@secondoftwo
 \fi
}%
\providecommand \@ifx [1]{%
 \ifx #1\expandafter \@firstoftwo
 \else \expandafter \@secondoftwo
 \fi
}%
\providecommand \natexlab [1]{#1}%
\providecommand \enquote  [1]{``#1''}%
\providecommand \bibnamefont  [1]{#1}%
\providecommand \bibfnamefont [1]{#1}%
\providecommand \citenamefont [1]{#1}%
\providecommand \href@noop [0]{\@secondoftwo}%
\providecommand \href [0]{\begingroup \@sanitize@url \@href}%
\providecommand \@href[1]{\@@startlink{#1}\@@href}%
\providecommand \@@href[1]{\endgroup#1\@@endlink}%
\providecommand \@sanitize@url [0]{\catcode `\\12\catcode `\$12\catcode
  `\&12\catcode `\#12\catcode `\^12\catcode `\_12\catcode `\%12\relax}%
\providecommand \@@startlink[1]{}%
\providecommand \@@endlink[0]{}%
\providecommand \url  [0]{\begingroup\@sanitize@url \@url }%
\providecommand \@url [1]{\endgroup\@href {#1}{\urlprefix }}%
\providecommand \urlprefix  [0]{URL }%
\providecommand \Eprint [0]{\href }%
\providecommand \doibase [0]{http://dx.doi.org/}%
\providecommand \selectlanguage [0]{\@gobble}%
\providecommand \bibinfo  [0]{\@secondoftwo}%
\providecommand \bibfield  [0]{\@secondoftwo}%
\providecommand \translation [1]{[#1]}%
\providecommand \BibitemOpen [0]{}%
\providecommand \bibitemStop [0]{}%
\providecommand \bibitemNoStop [0]{.\EOS\space}%
\providecommand \EOS [0]{\spacefactor3000\relax}%
\providecommand \BibitemShut  [1]{\csname bibitem#1\endcsname}%
\let\auto@bib@innerbib\@empty
%</preamble>
\bibitem [{\citenamefont {Karsch}(2002)}]{Karsch2002}%
  \BibitemOpen
  \bibfield  {author} {\bibinfo {author} {\bibfnamefont {F.}~\bibnamefont
  {Karsch}},\ }\bibfield  {booktitle} {\emph {\bibinfo {booktitle} {{Lectures
  on quark matter. Proceedings, 40. International Universit\"{a}tswochen for
  theoretical physics, 40th Winter School, IUKT 40}}},\ }\href {\doibase
  10.1007/3-540-45792-5_6} {\bibfield  {journal} {\bibinfo  {journal} {Lect.
  Notes Phys.}\ }\textbf {\bibinfo {volume} {583}},\ \bibinfo {pages} {209}
  (\bibinfo {year} {2002})},\ \Eprint {http://arxiv.org/abs/hep-lat/0106019}
  {arXiv:hep-lat/0106019 [hep-lat]} \BibitemShut {NoStop}%
\bibitem [{\citenamefont {Fukushima}\ and\ \citenamefont
  {Hatsuda}(2011)}]{Fukushima2011}%
  \BibitemOpen
  \bibfield  {author} {\bibinfo {author} {\bibfnamefont {K.}~\bibnamefont
  {Fukushima}}\ and\ \bibinfo {author} {\bibfnamefont {T.}~\bibnamefont
  {Hatsuda}},\ }\href {\doibase 10.1088/0034-4885/74/1/014001} {\bibfield
  {journal} {\bibinfo  {journal} {Rept. Prog. Phys.}\ }\textbf {\bibinfo
  {volume} {74}},\ \bibinfo {pages} {014001} (\bibinfo {year} {2011})},\
  \Eprint {http://arxiv.org/abs/1005.4814} {arXiv:1005.4814 [hep-ph]}
  \BibitemShut {NoStop}%
\bibitem [{\citenamefont {Gattringer}\ and\ \citenamefont
  {Langfeld}(2016)}]{Gattringer2016}%
  \BibitemOpen
  \bibfield  {author} {\bibinfo {author} {\bibfnamefont {C.}~\bibnamefont
  {Gattringer}}\ and\ \bibinfo {author} {\bibfnamefont {K.}~\bibnamefont
  {Langfeld}},\ }\href {\doibase 10.1142/S0217751X16430077} {\bibfield
  {journal} {\bibinfo  {journal} {Int. J. Mod. Phys.}\ }\textbf {\bibinfo
  {volume} {A31}},\ \bibinfo {pages} {1643007} (\bibinfo {year} {2016})},\
  \Eprint {http://arxiv.org/abs/1603.09517} {arXiv:1603.09517 [hep-lat]}
  \BibitemShut {NoStop}%
\bibitem [{\citenamefont {Fischer}(2006)}]{Fischer2006}%
  \BibitemOpen
  \bibfield  {author} {\bibinfo {author} {\bibfnamefont {C.~S.}\ \bibnamefont
  {Fischer}},\ }\href {\doibase 10.1088/0954-3899/32/8/R02} {\bibfield
  {journal} {\bibinfo  {journal} {J. Phys.}\ }\textbf {\bibinfo {volume}
  {G32}},\ \bibinfo {pages} {R253} (\bibinfo {year} {2006})},\ \Eprint
  {http://arxiv.org/abs/hep-ph/0605173} {arXiv:hep-ph/0605173 [hep-ph]}
  \BibitemShut {NoStop}%
\bibitem [{\citenamefont {Alkofer}\ and\ \citenamefont {von
  Smekal}(2001)}]{Alkofer2001}%
  \BibitemOpen
  \bibfield  {author} {\bibinfo {author} {\bibfnamefont {R.}~\bibnamefont
  {Alkofer}}\ and\ \bibinfo {author} {\bibfnamefont {L.}~\bibnamefont {von
  Smekal}},\ }\href {\doibase 10.1016/S0370-1573(01)00010-2} {\bibfield
  {journal} {\bibinfo  {journal} {Phys. Rept.}\ }\textbf {\bibinfo {volume}
  {353}},\ \bibinfo {pages} {281} (\bibinfo {year} {2001})},\ \Eprint
  {http://arxiv.org/abs/hep-ph/0007355} {arXiv:hep-ph/0007355 [hep-ph]}
  \BibitemShut {NoStop}%
\bibitem [{\citenamefont {Binosi}\ and\ \citenamefont
  {Papavassiliou}(2009)}]{Binosi2009}%
  \BibitemOpen
  \bibfield  {author} {\bibinfo {author} {\bibfnamefont {D.}~\bibnamefont
  {Binosi}}\ and\ \bibinfo {author} {\bibfnamefont {J.}~\bibnamefont
  {Papavassiliou}},\ }\href {\doibase 10.1016/j.physrep.2009.05.001} {\bibfield
   {journal} {\bibinfo  {journal} {Phys. Rept.}\ }\textbf {\bibinfo {volume}
  {479}},\ \bibinfo {pages} {1} (\bibinfo {year} {2009})},\ \Eprint
  {http://arxiv.org/abs/0909.2536} {arXiv:0909.2536 [hep-ph]} \BibitemShut
  {NoStop}%
\bibitem [{\citenamefont {Pawlowski}(2007)}]{Pawlowski2007}%
  \BibitemOpen
  \bibfield  {author} {\bibinfo {author} {\bibfnamefont {J.~M.}\ \bibnamefont
  {Pawlowski}},\ }\href {\doibase 10.1016/j.aop.2007.01.007} {\bibfield
  {journal} {\bibinfo  {journal} {Annals Phys.}\ }\textbf {\bibinfo {volume}
  {322}},\ \bibinfo {pages} {2831} (\bibinfo {year} {2007})},\ \Eprint
  {http://arxiv.org/abs/hep-th/0512261} {arXiv:hep-th/0512261 [hep-th]}
  \BibitemShut {NoStop}%
\bibitem [{\citenamefont {Gies}(2012)}]{Gies2012}%
  \BibitemOpen
  \bibfield  {author} {\bibinfo {author} {\bibfnamefont {H.}~\bibnamefont
  {Gies}},\ }\bibfield  {booktitle} {\emph {\bibinfo {booktitle} {{ECT* School
  on Renormalization Group and Effective Field Theory Approaches to Many-Body
  Systems Trento, Italy, February 27-March 10, 2006}}},\ }\href {\doibase
  10.1007/978-3-642-27320-9_6} {\bibfield  {journal} {\bibinfo  {journal}
  {Lect. Notes Phys.}\ }\textbf {\bibinfo {volume} {852}},\ \bibinfo {pages}
  {287} (\bibinfo {year} {2012})},\ \Eprint
  {http://arxiv.org/abs/hep-ph/0611146} {arXiv:hep-ph/0611146 [hep-ph]}
  \BibitemShut {NoStop}%
\bibitem [{\citenamefont {Quandt}\ \emph {et~al.}(2014)\citenamefont {Quandt},
  \citenamefont {Reinhardt},\ and\ \citenamefont {Heffner}}]{Quandt2014}%
  \BibitemOpen
  \bibfield  {author} {\bibinfo {author} {\bibfnamefont {M.}~\bibnamefont
  {Quandt}}, \bibinfo {author} {\bibfnamefont {H.}~\bibnamefont {Reinhardt}}, \
  and\ \bibinfo {author} {\bibfnamefont {J.}~\bibnamefont {Heffner}},\ }\href
  {\doibase 10.1103/PhysRevD.89.065037} {\bibfield  {journal} {\bibinfo
  {journal} {Phys. Rev. D}\ }\textbf {\bibinfo {volume} {89}},\ \bibinfo
  {pages} {065037} (\bibinfo {year} {2014})},\ \Eprint
  {http://arxiv.org/abs/1310.5950} {arXiv:1310.5950 [hep-th]} \BibitemShut
  {NoStop}%
\bibitem [{\citenamefont {Quandt}\ and\ \citenamefont
  {Reinhardt}(2015)}]{Quandt2015}%
  \BibitemOpen
  \bibfield  {author} {\bibinfo {author} {\bibfnamefont {M.}~\bibnamefont
  {Quandt}}\ and\ \bibinfo {author} {\bibfnamefont {H.}~\bibnamefont
  {Reinhardt}},\ }\href {\doibase 10.1103/PhysRevD.92.025051} {\bibfield
  {journal} {\bibinfo  {journal} {Phys. Rev. D}\ }\textbf {\bibinfo {volume}
  {92}},\ \bibinfo {pages} {025051} (\bibinfo {year} {2015})},\ \Eprint
  {http://arxiv.org/abs/1503.06993} {arXiv:1503.06993 [hep-th]} \BibitemShut
  {NoStop}%
\bibitem [{\citenamefont {Feuchter}\ and\ \citenamefont
  {Reinhardt}(2004)}]{Feuchter2004}%
  \BibitemOpen
  \bibfield  {author} {\bibinfo {author} {\bibfnamefont {C.}~\bibnamefont
  {Feuchter}}\ and\ \bibinfo {author} {\bibfnamefont {H.}~\bibnamefont
  {Reinhardt}},\ }\href {\doibase 10.1103/PhysRevD.70.105021} {\bibfield
  {journal} {\bibinfo  {journal} {Phys. Rev. D}\ }\textbf {\bibinfo {volume}
  {70}},\ \bibinfo {pages} {105021} (\bibinfo {year} {2004})},\ \Eprint
  {http://arxiv.org/abs/hep-th/0408236} {arXiv:hep-th/0408236} \BibitemShut
  {NoStop}%
\bibitem [{\citenamefont {Reinhardt}\ and\ \citenamefont
  {Feuchter}(2005)}]{Feuchter2005}%
  \BibitemOpen
  \bibfield  {author} {\bibinfo {author} {\bibfnamefont {H.}~\bibnamefont
  {Reinhardt}}\ and\ \bibinfo {author} {\bibfnamefont {C.}~\bibnamefont
  {Feuchter}},\ }\href@noop {} {\bibfield  {journal} {\bibinfo  {journal}
  {Phys. Rev. D}\ }\textbf {\bibinfo {volume} {71}},\ \bibinfo {pages} {105002}
  (\bibinfo {year} {2005})},\ \Eprint {http://arxiv.org/abs/hep-th/0408237}
  {arXiv:hep-th/0408237} \BibitemShut {NoStop}%
\bibitem [{\citenamefont {Reinhardt}\ \emph {et~al.}(2017)\citenamefont
  {Reinhardt}, \citenamefont {Burgio}, \citenamefont {Campagnari},
  \citenamefont {Ebadati}, \citenamefont {Heffner}, \citenamefont {Quandt},
  \citenamefont {Vastag},\ and\ \citenamefont {Vogt}}]{Reinhardt2017}%
  \BibitemOpen
  \bibfield  {author} {\bibinfo {author} {\bibfnamefont {H.}~\bibnamefont
  {Reinhardt}}, \bibinfo {author} {\bibfnamefont {G.}~\bibnamefont {Burgio}},
  \bibinfo {author} {\bibfnamefont {D.}~\bibnamefont {Campagnari}}, \bibinfo
  {author} {\bibfnamefont {E.}~\bibnamefont {Ebadati}}, \bibinfo {author}
  {\bibfnamefont {J.}~\bibnamefont {Heffner}}, \bibinfo {author} {\bibfnamefont
  {M.}~\bibnamefont {Quandt}}, \bibinfo {author} {\bibfnamefont
  {P.}~\bibnamefont {Vastag}}, \ and\ \bibinfo {author} {\bibfnamefont
  {H.}~\bibnamefont {Vogt}}\ }(\bibinfo {year} {2017})\ \Eprint
  {http://arxiv.org/abs/1706.02702} {arXiv:1706.02702 [hep-th]} \BibitemShut
  {NoStop}%
\bibitem [{\citenamefont {Reinhardt}\ and\ \citenamefont
  {Vastag}(2016)}]{RV2016}%
  \BibitemOpen
  \bibfield  {author} {\bibinfo {author} {\bibfnamefont {H.}~\bibnamefont
  {Reinhardt}}\ and\ \bibinfo {author} {\bibfnamefont {P.}~\bibnamefont
  {Vastag}},\ }\href {\doibase 10.1103/PhysRevD.94.105005} {\bibfield
  {journal} {\bibinfo  {journal} {Phys. Rev. D}\ }\textbf {\bibinfo {volume}
  {94}},\ \bibinfo {pages} {105005} (\bibinfo {year} {2016})},\ \Eprint
  {http://arxiv.org/abs/1605.03740} {arXiv:1605.03740 [hep-th]} \BibitemShut
  {NoStop}%
\bibitem [{\citenamefont {Reinhardt}(2016)}]{Reinhardt2016}%
  \BibitemOpen
  \bibfield  {author} {\bibinfo {author} {\bibfnamefont {H.}~\bibnamefont
  {Reinhardt}},\ }\href {\doibase 10.1103/PhysRevD.94.045016} {\bibfield
  {journal} {\bibinfo  {journal} {Phys. Rev. D}\ }\textbf {\bibinfo {volume}
  {94}},\ \bibinfo {pages} {045016} (\bibinfo {year} {2016})},\ \Eprint
  {http://arxiv.org/abs/1604.06273} {arXiv:1604.06273 [hep-th]} \BibitemShut
  {NoStop}%
\bibitem [{\citenamefont {Adler}\ and\ \citenamefont
  {Davis}(1984)}]{Adler1984}%
  \BibitemOpen
  \bibfield  {author} {\bibinfo {author} {\bibfnamefont {S.}~\bibnamefont
  {Adler}}\ and\ \bibinfo {author} {\bibfnamefont {A.}~\bibnamefont {Davis}},\
  }\href@noop {} {\bibfield  {journal} {\bibinfo  {journal} {Nuclear Physics
  B}\ }\textbf {\bibinfo {volume} {244}},\ \bibinfo {pages} {469} (\bibinfo
  {year} {1984})}\BibitemShut {NoStop}%
\bibitem [{\citenamefont {Davis}\ and\ \citenamefont
  {Matheson}(1984)}]{Davis1984}%
  \BibitemOpen
  \bibfield  {author} {\bibinfo {author} {\bibfnamefont {A.}~\bibnamefont
  {Davis}}\ and\ \bibinfo {author} {\bibfnamefont {A.}~\bibnamefont
  {Matheson}},\ }\href {\doibase 10.1016/0550-3213(84)90292-X} {\bibfield
  {journal} {\bibinfo  {journal} {Nuclear Physics B}\ }\textbf {\bibinfo
  {volume} {246}},\ \bibinfo {pages} {203} (\bibinfo {year}
  {1984})}\BibitemShut {NoStop}%
\bibitem [{\citenamefont {Koci\'{c}}(1986)}]{Kocic1986}%
  \BibitemOpen
  \bibfield  {author} {\bibinfo {author} {\bibfnamefont {A.}~\bibnamefont
  {Koci\'{c}}},\ }\href {\doibase 10.1103/PhysRevD.33.1785} {\bibfield
  {journal} {\bibinfo  {journal} {Phys. Rev. D}\ }\textbf {\bibinfo {volume}
  {33}},\ \bibinfo {pages} {1785} (\bibinfo {year} {1986})}\BibitemShut
  {NoStop}%
\bibitem [{\citenamefont {Alkofer}\ \emph {et~al.}(1989)\citenamefont
  {Alkofer}, \citenamefont {Amundsen},\ and\ \citenamefont
  {Langfeld}}]{Alkofer1989}%
  \BibitemOpen
  \bibfield  {author} {\bibinfo {author} {\bibfnamefont {R.}~\bibnamefont
  {Alkofer}}, \bibinfo {author} {\bibfnamefont {P.~A.}\ \bibnamefont
  {Amundsen}}, \ and\ \bibinfo {author} {\bibfnamefont {K.}~\bibnamefont
  {Langfeld}},\ }\href {\doibase 10.1007/BF01555857} {\bibfield  {journal}
  {\bibinfo  {journal} {Z. Phys.}\ }\textbf {\bibinfo {volume} {C42}},\
  \bibinfo {pages} {199} (\bibinfo {year} {1989})}\BibitemShut {NoStop}%
\bibitem [{\citenamefont {Lo}\ and\ \citenamefont {Swanson}(2010)}]{LS2010}%
  \BibitemOpen
  \bibfield  {author} {\bibinfo {author} {\bibfnamefont {P.~M.}\ \bibnamefont
  {Lo}}\ and\ \bibinfo {author} {\bibfnamefont {E.~S.}\ \bibnamefont
  {Swanson}},\ }\href {\doibase 10.1103/PhysRevD.81.034030} {\bibfield
  {journal} {\bibinfo  {journal} {Phys. Rev. D}\ }\textbf {\bibinfo {volume}
  {81}},\ \bibinfo {pages} {034030} (\bibinfo {year} {2010})},\ \Eprint
  {http://arxiv.org/abs/0908.4099} {arXiv:0908.4099 [hep-ph]} \BibitemShut
  {NoStop}%
\bibitem [{\citenamefont {Reinhardt}\ and\ \citenamefont
  {Heffner}(2013)}]{RH2013}%
  \BibitemOpen
  \bibfield  {author} {\bibinfo {author} {\bibfnamefont {H.}~\bibnamefont
  {Reinhardt}}\ and\ \bibinfo {author} {\bibfnamefont {J.}~\bibnamefont
  {Heffner}},\ }\href {\doibase 10.1103/PhysRevD.88.045024} {\bibfield
  {journal} {\bibinfo  {journal} {Phys. Rev. D}\ }\textbf {\bibinfo {volume}
  {88}},\ \bibinfo {pages} {045024} (\bibinfo {year} {2013})},\ \Eprint
  {http://arxiv.org/abs/1304.2980} {arXiv:1304.2980} \BibitemShut {NoStop}%
\bibitem [{\citenamefont {Christ}\ and\ \citenamefont
  {Lee}(1980)}]{Christ1980}%
  \BibitemOpen
  \bibfield  {author} {\bibinfo {author} {\bibfnamefont {N.~H.}\ \bibnamefont
  {Christ}}\ and\ \bibinfo {author} {\bibfnamefont {T.~D.}\ \bibnamefont
  {Lee}},\ }\href@noop {} {\bibfield  {journal} {\bibinfo  {journal} {Phys.
  Rev. D}\ }\textbf {\bibinfo {volume} {22}},\ \bibinfo {pages} {939} (\bibinfo
  {year} {1980})}\BibitemShut {NoStop}%
\bibitem [{\citenamefont {Heffner}\ and\ \citenamefont
  {Reinhardt}(2015)}]{Heffner2015}%
  \BibitemOpen
  \bibfield  {author} {\bibinfo {author} {\bibfnamefont {J.}~\bibnamefont
  {Heffner}}\ and\ \bibinfo {author} {\bibfnamefont {H.}~\bibnamefont
  {Reinhardt}},\ }\href {\doibase 10.1103/PhysRevD.91.085022} {\bibfield
  {journal} {\bibinfo  {journal} {Phys. Rev. D}\ }\textbf {\bibinfo {volume}
  {91}},\ \bibinfo {pages} {085022} (\bibinfo {year} {2015})},\ \Eprint
  {http://arxiv.org/abs/1501.05858} {arXiv:1501.05858} \BibitemShut {NoStop}%
\bibitem [{\citenamefont {Campagnari}\ \emph {et~al.}(2016)\citenamefont
  {Campagnari}, \citenamefont {Ebadati}, \citenamefont {Reinhardt},\ and\
  \citenamefont {Vastag}}]{QCDT0Rev}%
  \BibitemOpen
  \bibfield  {author} {\bibinfo {author} {\bibfnamefont {D.~R.}\ \bibnamefont
  {Campagnari}}, \bibinfo {author} {\bibfnamefont {E.}~\bibnamefont {Ebadati}},
  \bibinfo {author} {\bibfnamefont {H.}~\bibnamefont {Reinhardt}}, \ and\
  \bibinfo {author} {\bibfnamefont {P.}~\bibnamefont {Vastag}},\ }\href
  {\doibase 10.1103/PhysRevD.94.074027} {\bibfield  {journal} {\bibinfo
  {journal} {Phys. Rev. D}\ }\textbf {\bibinfo {volume} {94}},\ \bibinfo
  {pages} {074027} (\bibinfo {year} {2016})},\ \Eprint
  {http://arxiv.org/abs/1608.06820} {arXiv:1608.06820 [hep-ph]} \BibitemShut
  {NoStop}%
\bibitem [{\citenamefont {Epple}\ \emph {et~al.}(2007)\citenamefont {Epple},
  \citenamefont {Reinhardt},\ and\ \citenamefont {Schleifenbaum}}]{ERS2007}%
  \BibitemOpen
  \bibfield  {author} {\bibinfo {author} {\bibfnamefont {D.}~\bibnamefont
  {Epple}}, \bibinfo {author} {\bibfnamefont {H.}~\bibnamefont {Reinhardt}}, \
  and\ \bibinfo {author} {\bibfnamefont {W.}~\bibnamefont {Schleifenbaum}},\
  }\href@noop {} {\bibfield  {journal} {\bibinfo  {journal} {Phys. Rev. D}\
  }\textbf {\bibinfo {volume} {75}},\ \bibinfo {pages} {045011} (\bibinfo
  {year} {2007})},\ \Eprint {http://arxiv.org/abs/hep-th/0612241}
  {arXiv:hep-th/0612241} \BibitemShut {NoStop}%
\bibitem [{\citenamefont {Finger}\ and\ \citenamefont
  {Mandula}(1982)}]{FM1982}%
  \BibitemOpen
  \bibfield  {author} {\bibinfo {author} {\bibfnamefont {J.~R.}\ \bibnamefont
  {Finger}}\ and\ \bibinfo {author} {\bibfnamefont {J.~E.}\ \bibnamefont
  {Mandula}},\ }\href {\doibase 10.1016/0550-3213(82)90570-3} {\bibfield
  {journal} {\bibinfo  {journal} {Nuclear Physics B}\ }\textbf {\bibinfo
  {volume} {199}},\ \bibinfo {pages} {168} (\bibinfo {year}
  {1982})}\BibitemShut {NoStop}%
\bibitem [{\citenamefont {{Le Yaouanc}}\ \emph {et~al.}(1984)\citenamefont {{Le
  Yaouanc}}, \citenamefont {Oliver}, \citenamefont {P\`{e}ne},\ and\
  \citenamefont {Raynal}}]{LeYaouanc1984}%
  \BibitemOpen
  \bibfield  {author} {\bibinfo {author} {\bibfnamefont {A.}~\bibnamefont {{Le
  Yaouanc}}}, \bibinfo {author} {\bibfnamefont {L.}~\bibnamefont {Oliver}},
  \bibinfo {author} {\bibfnamefont {O.}~\bibnamefont {P\`{e}ne}}, \ and\
  \bibinfo {author} {\bibfnamefont {J.-C.}\ \bibnamefont {Raynal}},\ }\href
  {\doibase 10.1103/PhysRevD.29.1233} {\bibfield  {journal} {\bibinfo
  {journal} {Phys. Rev. D}\ }\textbf {\bibinfo {volume} {29}},\ \bibinfo
  {pages} {1233} (\bibinfo {year} {1984})}\BibitemShut {NoStop}%
\bibitem [{\citenamefont {Alkofer}\ and\ \citenamefont
  {Amundsen}(1988)}]{AA1988}%
  \BibitemOpen
  \bibfield  {author} {\bibinfo {author} {\bibfnamefont {R.}~\bibnamefont
  {Alkofer}}\ and\ \bibinfo {author} {\bibfnamefont {P.}~\bibnamefont
  {Amundsen}},\ }\href {\doibase 10.1016/0550-3213(88)90695-5} {\bibfield
  {journal} {\bibinfo  {journal} {Nuclear Physics B}\ }\textbf {\bibinfo
  {volume} {306}},\ \bibinfo {pages} {305} (\bibinfo {year}
  {1988})}\BibitemShut {NoStop}%
\bibitem [{\citenamefont {Alkofer}\ and\ \citenamefont
  {Amundsen}(1987)}]{Alkofer1986}%
  \BibitemOpen
  \bibfield  {author} {\bibinfo {author} {\bibfnamefont {R.}~\bibnamefont
  {Alkofer}}\ and\ \bibinfo {author} {\bibfnamefont {P.~A.}\ \bibnamefont
  {Amundsen}},\ }\href {\doibase 10.1016/0370-2693(87)91117-8} {\bibfield
  {journal} {\bibinfo  {journal} {Phys. Lett.}\ }\textbf {\bibinfo {volume}
  {B187}},\ \bibinfo {pages} {395} (\bibinfo {year} {1987})}\BibitemShut
  {NoStop}%
\bibitem [{\citenamefont {Campagnari}\ \emph {et~al.}(2009)\citenamefont
  {Campagnari}, \citenamefont {Reinhardt},\ and\ \citenamefont
  {Weber}}]{Campagnari2009}%
  \BibitemOpen
  \bibfield  {author} {\bibinfo {author} {\bibfnamefont {D.~R.}\ \bibnamefont
  {Campagnari}}, \bibinfo {author} {\bibfnamefont {H.}~\bibnamefont
  {Reinhardt}}, \ and\ \bibinfo {author} {\bibfnamefont {A.}~\bibnamefont
  {Weber}},\ }\href {\doibase 10.1103/PhysRevD.80.025005} {\bibfield  {journal}
  {\bibinfo  {journal} {Phys. Rev. D}\ }\textbf {\bibinfo {volume} {80}},\
  \bibinfo {pages} {025005} (\bibinfo {year} {2009})},\ \Eprint
  {http://arxiv.org/abs/0904.3490} {arXiv:0904.3490 [hep-th]} \BibitemShut
  {NoStop}%
\bibitem [{\citenamefont {Nakagawa}\ \emph {et~al.}(2006)\citenamefont
  {Nakagawa}, \citenamefont {Nakamura}, \citenamefont {Saito}, \citenamefont
  {Toki},\ and\ \citenamefont {Zwanziger}}]{Nakagawa:2006fk}%
  \BibitemOpen
  \bibfield  {author} {\bibinfo {author} {\bibfnamefont {Y.}~\bibnamefont
  {Nakagawa}}, \bibinfo {author} {\bibfnamefont {A.}~\bibnamefont {Nakamura}},
  \bibinfo {author} {\bibfnamefont {T.}~\bibnamefont {Saito}}, \bibinfo
  {author} {\bibfnamefont {H.}~\bibnamefont {Toki}}, \ and\ \bibinfo {author}
  {\bibfnamefont {D.}~\bibnamefont {Zwanziger}},\ }\href {\doibase
  10.1103/PhysRevD.73.094504} {\bibfield  {journal} {\bibinfo  {journal} {Phys.
  Rev.}\ }\textbf {\bibinfo {volume} {D73}},\ \bibinfo {pages} {094504}
  (\bibinfo {year} {2006})},\ \Eprint {http://arxiv.org/abs/hep-lat/0603010}
  {arXiv:hep-lat/0603010 [hep-lat]} \BibitemShut {NoStop}%
%%CITATION = HEP-LAT/0603010;%%
\bibitem [{\citenamefont {Greensite}\ and\ \citenamefont
  {Szczepaniak}(2015)}]{Greensite:2014bua}%
  \BibitemOpen
  \bibfield  {author} {\bibinfo {author} {\bibfnamefont {J.}~\bibnamefont
  {Greensite}}\ and\ \bibinfo {author} {\bibfnamefont {A.~P.}\ \bibnamefont
  {Szczepaniak}},\ }\href {\doibase 10.1103/PhysRevD.91.034503} {\bibfield
  {journal} {\bibinfo  {journal} {Phys. Rev.}\ }\textbf {\bibinfo {volume}
  {D91}},\ \bibinfo {pages} {034503} (\bibinfo {year} {2015})},\ \Eprint
  {http://arxiv.org/abs/1410.3525} {arXiv:1410.3525 [hep-lat]} \BibitemShut
  {NoStop}%
%%CITATION = ARXIV:1410.3525;%%
\bibitem [{\citenamefont {Golterman}\ \emph {et~al.}(2012)\citenamefont
  {Golterman}, \citenamefont {Greensite}, \citenamefont {Peris},\ and\
  \citenamefont {Szczepaniak}}]{Golterman:2012dx}%
  \BibitemOpen
  \bibfield  {author} {\bibinfo {author} {\bibfnamefont {M.}~\bibnamefont
  {Golterman}}, \bibinfo {author} {\bibfnamefont {J.}~\bibnamefont
  {Greensite}}, \bibinfo {author} {\bibfnamefont {S.}~\bibnamefont {Peris}}, \
  and\ \bibinfo {author} {\bibfnamefont {A.~P.}\ \bibnamefont {Szczepaniak}},\
  }\href {\doibase 10.1103/PhysRevD.85.085016} {\bibfield  {journal} {\bibinfo
  {journal} {Phys. Rev.}\ }\textbf {\bibinfo {volume} {D85}},\ \bibinfo {pages}
  {085016} (\bibinfo {year} {2012})},\ \Eprint {http://arxiv.org/abs/1201.4590}
  {arXiv:1201.4590 [hep-th]} \BibitemShut {NoStop}%
%%CITATION = ARXIV:1201.4590;%%
\bibitem [{\citenamefont {Burgio}\ \emph {et~al.}(2015)\citenamefont {Burgio},
  \citenamefont {Quandt}, \citenamefont {Reinhardt},\ and\ \citenamefont
  {Vogt}}]{Burgio:2015hsa}%
  \BibitemOpen
  \bibfield  {author} {\bibinfo {author} {\bibfnamefont {G.}~\bibnamefont
  {Burgio}}, \bibinfo {author} {\bibfnamefont {M.}~\bibnamefont {Quandt}},
  \bibinfo {author} {\bibfnamefont {H.}~\bibnamefont {Reinhardt}}, \ and\
  \bibinfo {author} {\bibfnamefont {H.}~\bibnamefont {Vogt}},\ }\href {\doibase
  10.1103/PhysRevD.92.034518} {\bibfield  {journal} {\bibinfo  {journal} {Phys.
  Rev.}\ }\textbf {\bibinfo {volume} {D92}},\ \bibinfo {pages} {034518}
  (\bibinfo {year} {2015})},\ \Eprint {http://arxiv.org/abs/1503.09064}
  {arXiv:1503.09064 [hep-lat]} \BibitemShut {NoStop}%
%%CITATION = ARXIV:1503.09064;%%
\bibitem [{\citenamefont {Burgio}\ \emph {et~al.}(2017)\citenamefont {Burgio},
  \citenamefont {Quandt}, \citenamefont {Reinhardt},\ and\ \citenamefont
  {Vogt}}]{Burgio:2016nad}%
  \BibitemOpen
  \bibfield  {author} {\bibinfo {author} {\bibfnamefont {G.}~\bibnamefont
  {Burgio}}, \bibinfo {author} {\bibfnamefont {M.}~\bibnamefont {Quandt}},
  \bibinfo {author} {\bibfnamefont {H.}~\bibnamefont {Reinhardt}}, \ and\
  \bibinfo {author} {\bibfnamefont {H.}~\bibnamefont {Vogt}},\ }\href {\doibase
  10.1103/PhysRevD.95.014503} {\bibfield  {journal} {\bibinfo  {journal} {Phys.
  Rev.}\ }\textbf {\bibinfo {volume} {D95}},\ \bibinfo {pages} {014503}
  (\bibinfo {year} {2017})},\ \Eprint {http://arxiv.org/abs/1608.05795}
  {arXiv:1608.05795 [hep-lat]} \BibitemShut {NoStop}%
%%CITATION = ARXIV:1608.05795;%%
\bibitem [{\citenamefont {Aitken}(1926)}]{Aitken:1926}%
  \BibitemOpen
  \bibfield  {author} {\bibinfo {author} {\bibfnamefont {A.}~\bibnamefont
  {Aitken}},\ }\href@noop {} {\bibfield  {journal} {\bibinfo  {journal}
  {Proc.~Roy.~Soc.~Edinburgh}\ }\textbf {\bibinfo {volume} {A46}},\ \bibinfo
  {pages} {289} (\bibinfo {year} {1926})}\BibitemShut {NoStop}%
\bibitem [{\citenamefont {Anderson}(1965)}]{Anderson:1965}%
  \BibitemOpen
  \bibfield  {author} {\bibinfo {author} {\bibfnamefont {D.}~\bibnamefont
  {Anderson}},\ }\href@noop {} {\bibfield  {journal} {\bibinfo  {journal}
  {Journal ACM}\ }\textbf {\bibinfo {volume} {12}},\ \bibinfo {pages} {547}
  (\bibinfo {year} {1965})}\BibitemShut {NoStop}%
\bibitem [{\citenamefont {Campagnari}\ and\ \citenamefont
  {Reinhardt}(2018)}]{Campagnari:2018flz}%
  \BibitemOpen
  \bibfield  {author} {\bibinfo {author} {\bibfnamefont {D.}~\bibnamefont
  {Campagnari}}\ and\ \bibinfo {author} {\bibfnamefont {H.}~\bibnamefont
  {Reinhardt}},\ }\href {\doibase 10.1103/PhysRevD.97.054027} {\bibfield
  {journal} {\bibinfo  {journal} {Phys. Rev.}\ }\textbf {\bibinfo {volume}
  {D97}},\ \bibinfo {pages} {054027} (\bibinfo {year} {2018})},\ \Eprint
  {http://arxiv.org/abs/1801.02045} {arXiv:1801.02045 [hep-th]} \BibitemShut
  {NoStop}%
%%CITATION = ARXIV:1801.02045;%%
\bibitem [{\citenamefont {Wynn}(1962)}]{Wynn:1952}%
  \BibitemOpen
  \bibfield  {author} {\bibinfo {author} {\bibfnamefont {P.}~\bibnamefont
  {Wynn}},\ }\href@noop {} {\bibfield  {journal} {\bibinfo  {journal}
  {Math.~Comp.}\ }\textbf {\bibinfo {volume} {16}},\ \bibinfo {pages} {301}
  (\bibinfo {year} {1962})}\BibitemShut {NoStop}%
\bibitem [{\citenamefont {Wynn}(1956)}]{Wynn:1956}%
  \BibitemOpen
  \bibfield  {author} {\bibinfo {author} {\bibfnamefont {P.}~\bibnamefont
  {Wynn}},\ }\href@noop {} {\bibfield  {journal} {\bibinfo  {journal}
  {Math.~Tables Automat.~Comp.}\ }\textbf {\bibinfo {volume} {10}},\ \bibinfo
  {pages} {91} (\bibinfo {year} {1956})}\BibitemShut {NoStop}%
\bibitem [{\citenamefont {Greensite}\ and\ \citenamefont
  {Olejnik}(2003)}]{Greensite2003}%
  \BibitemOpen
  \bibfield  {author} {\bibinfo {author} {\bibfnamefont {J.}~\bibnamefont
  {Greensite}}\ and\ \bibinfo {author} {\bibfnamefont {S.}~\bibnamefont
  {Olejnik}},\ }\href {\doibase 10.1103/PhysRevD.67.094503} {\bibfield
  {journal} {\bibinfo  {journal} {Phys. Rev. D}\ }\textbf {\bibinfo {volume}
  {67}},\ \bibinfo {pages} {094503} (\bibinfo {year} {2003})},\ \Eprint
  {http://arxiv.org/abs/hep-lat/0302018} {arXiv:hep-lat/0302018 [hep-lat]}
  \BibitemShut {NoStop}%
\bibitem [{\citenamefont {Bors\'{a}nyi}\ \emph {et~al.}(2010)\citenamefont
  {Bors\'{a}nyi}, \citenamefont {Fodor}, \citenamefont {Hoelbling},
  \citenamefont {Katz}, \citenamefont {Krieg}, \citenamefont {Ratti},\ and\
  \citenamefont {Szab\'{o}}}]{Borsanyi2010}%
  \BibitemOpen
  \bibfield  {author} {\bibinfo {author} {\bibfnamefont {S.}~\bibnamefont
  {Bors\'{a}nyi}}, \bibinfo {author} {\bibfnamefont {Z.}~\bibnamefont {Fodor}},
  \bibinfo {author} {\bibfnamefont {C.}~\bibnamefont {Hoelbling}}, \bibinfo
  {author} {\bibfnamefont {S.~D.}\ \bibnamefont {Katz}}, \bibinfo {author}
  {\bibfnamefont {S.}~\bibnamefont {Krieg}}, \bibinfo {author} {\bibfnamefont
  {C.}~\bibnamefont {Ratti}}, \ and\ \bibinfo {author} {\bibfnamefont {K.~K.}\
  \bibnamefont {Szab\'{o}}} (\bibinfo {collaboration} {Wuppertal-Budapest}),\
  }\href {\doibase 10.1007/JHEP09(2010)073} {\bibfield  {journal} {\bibinfo
  {journal} {JHEP}\ }\textbf {\bibinfo {volume} {09}},\ \bibinfo {pages} {073}
  (\bibinfo {year} {2010})},\ \Eprint {http://arxiv.org/abs/1005.3508}
  {arXiv:1005.3508 [hep-lat]} \BibitemShut {NoStop}%
\bibitem [{\citenamefont {Bazavov}\ \emph {et~al.}(2012)\citenamefont {Bazavov}
  \emph {et~al.}}]{Bazavov2012}%
  \BibitemOpen
  \bibfield  {author} {\bibinfo {author} {\bibfnamefont {A.}~\bibnamefont
  {Bazavov}} \emph {et~al.},\ }\href {\doibase 10.1103/PhysRevD.85.054503}
  {\bibfield  {journal} {\bibinfo  {journal} {Phys. Rev. D}\ }\textbf {\bibinfo
  {volume} {85}},\ \bibinfo {pages} {054503} (\bibinfo {year} {2012})},\
  \Eprint {http://arxiv.org/abs/1111.1710} {arXiv:1111.1710 [hep-lat]}
  \BibitemShut {NoStop}%
\bibitem [{\citenamefont {Alkofer}\ \emph {et~al.}(2006)\citenamefont
  {Alkofer}, \citenamefont {Kloker}, \citenamefont {Krassnigg},\ and\
  \citenamefont {Wagenbrunn}}]{Alkofer2005}%
  \BibitemOpen
  \bibfield  {author} {\bibinfo {author} {\bibfnamefont {R.}~\bibnamefont
  {Alkofer}}, \bibinfo {author} {\bibfnamefont {M.}~\bibnamefont {Kloker}},
  \bibinfo {author} {\bibfnamefont {A.}~\bibnamefont {Krassnigg}}, \ and\
  \bibinfo {author} {\bibfnamefont {R.~F.}\ \bibnamefont {Wagenbrunn}},\ }\href
  {\doibase 10.1103/PhysRevLett.96.022001} {\bibfield  {journal} {\bibinfo
  {journal} {Phys. Rev. Lett.}\ }\textbf {\bibinfo {volume} {96}},\ \bibinfo
  {pages} {022001} (\bibinfo {year} {2006})},\ \Eprint
  {http://arxiv.org/abs/hep-ph/0510028} {arXiv:hep-ph/0510028 [hep-ph]}
  \BibitemShut {NoStop}%
\end{thebibliography}%

\end{document}